# Valley resolved dynamics of phonon bottleneck in semiconductor molybdenum ditelluride

Zhong Wang[1,2], Yijie Shi[1,2], Yu Pan[3], Min Li[4], Xi Wang[1,2], Zheng Zhang[1,2], Xiangyu Zhu[1,2], Fuyong Hua[1,2], Qian You[1,2,†], Chunlong Hu[1, 2], Junjie He[4], Yu Ye[3], Wenxi Liang[1,2,*]

[1]Wuhan National Laboratory for Optoelectronics, Huazhong University of Science and Technology, 1037 Luoyu Road, Wuhan 430074, China

[2]Advanced Biomedical Imaging Facility, Huazhong University of Science and Technology, Wuhan 430074, China

[3]State Key Laboratory for Mesoscopic Physics and Frontiers Science Center for Nano-Optoelectronics, School of Physics, Peking University, Beijing 100871, China

[4]Department of Physical and Macromolecular Chemistry, Faculty of Science, Charles University, Prague 12843, Czech Republic

[†]Current address: Zhongshan Lighting Fast Intellectual Property Rights Service Center, Zhongshan 528421, China

[*]Email: wxliang@hust.edu.cn

**Abstract**: Semiconductor molybdenum ditelluride (2H-$MoTe_2$) possess multiple valleys in the band structure, enriching its physical properties and potentials in applications. The understanding of its multivalley nature of fundamental processes involving population and relaxation of carriers and phonons is still evolving; particularly, the possible phonon bottleneck has not yet been addressed. Here, we investigate the carrier intra- and intervalley scattering and the phonon dynamics in different valleys in photoexcited few-layer 2H-$MoTe_2$, by using the time resolved measurements of optical absorption and electron diffraction, together with the density functional theory calculation and molecular dynamics simulation. The pathways and timescales of carrier relaxation, accompanied with the emissions of optical phonons at the Brillouin zone center and acoustic phonons at the zone border are revealed. We present a couple of approaches to estimate the population of different phonon modes based on the results of optical and electron diffraction measurements, hence quantitatively identify the occurrences of phonon bottleneck located in different valleys. Our findings make possible to construct a comprehensive picture of the complex interactions between carriers and phonons in 2H-$MoTe_2$ with the valley degree of freedom resolved.

## Introduction

Molybdenum ditelluride ($MoTe_2$) with three crystal phases[1] of semiconducting 2H, semimetallic 1T′, and type-II Weyl semimetallic $T_d$, offers an ideal platform material for studying the attractive family of transition metal dichalcogenides (TMDCs), which possesses abundant physical phenomena of, e.g., charge density wave[2], superconductivity[3], and topological electronic phase[4]. In terms of applicability, 2H-$MoTe_2$ is potentially compatible with the matured silicon-based industry[5] due to its tunable bandgap of ~1 eV, close to that of silicon, and has been implemented in various applications,[6] including the third generation solar cells[7], valleytronic devices[8], highly sensitive photodetectors[9], and phase-change memories[10]. The carrier and lattice dynamics of 2H-$MoTe_2$, which link the properties to the physical phenomena and device performances, have been studied through multiple time-resolved approaches[7,11-14]. But the complexity of microscale processes involving time spans, pathways, and interactions induced by the valley degree of freedom is yet to be elucidated, as the excited carriers in TMDCs evolves under the basic frame of intra- and intervalley scattering[15].

For the carrier relaxation in 2H-$MoTe_2$, the exciton dynamics have been extensively investigated, yielding quantitative results, e.g., the carrier multiplication in hundreds of femtoseconds[7] and the defect-assisted nonradiative recombination of band-edge excitons in a few picoseconds[14]. Optical probes are currently the main approaches to examine carrier dynamics, but the information obtained so far are inadequate to provide a comprehensive understanding for clarifying the mechanisms of electron-phonon scattering during non-radiative energy transfer[16], and especially for achieving a consensus on the conflicting observations. For example, the observations of core-level spectroscopy[13] and x-ray spectroscopy[12] found that the hot carriers, which lay critical impacts on photovoltaics, relax to the band edge within hundreds of femtoseconds, but the details of cooling path are still ambiguous. The study of lattice dynamics showed that the hot and band-edge electrons preferentially coupled to optical (OPs) and acoustic phonons (APs)[11], respectively, but the latter is inconsistent with the observed $A_{1g}$ phonon under infrared excitation[17]. The reported strong electron-phonon coupling[18] probably conflicts with some high order coupling within carriers, such as the impact ionization, contradicting to the high conversion efficiency of carrier multiplication[7]. Among all complex dynamics, the carrier migration path within multiple valleys and the role of different phonon modes involved are of particular interests, due to the hot topic of valleytronics in developing novel photoelectronics. To resolve the dynamics with multiple valleys in the band structure, the methodology with momentum sensitivity is demanded.

In this work, we investigate the energy relaxation carried by carriers and phonons in monocrystalline 2H-$MoTe_2$, through the combination of pump-probe based transient absorption (TA) spectroscopy and ultrafast electron diffraction (UED), and the density functional theory calculation and non-adiabatic molecular dynamics (NAMD) simulation. The relaxations of hot carriers and band-edge excitons, and the population and decoherence of OPs revealed by TA and NAMD, together with the momentum-resolved evolutions of strongly coupling phonon (SCP) observed by UED, construct the scenario of intra- and intervalley carrier scattering with dominating OPs at the Brillouin zone (BZ) center and APs at the BZ border, followed by the rapid lattice disorder before the completion of thermalization. With the computations of phonon population based on the results of time-resolved measurements, the phonon bottlenecks retarding the cooling processes of hot carriers are quantitatively identified in different valleys.

## Results

The 2H-$MoTe_2$ films with thickness of 6–7 nm (9–10 atomic layers) were epitaxially grown on a silicon wafer (see Methods)[19], then transferred onto bare grid as the free-standing (FS) specimen, and onto carbon membrane coated grid as the carbon-membrane-supported (CS) specimen with a heterostructure interface, as shown in Fig. 1a. The pump-probe setup of TA and UED is schematically illustrated in Supplementary Fig. S2. The hexagonal lattice structure and the diffraction pattern with four families of crystallographic plane recorded by UED are shown in Fig. 1a and 1b, respectively. We computed the electronic band structure and the electronic density of states using Vienna *Ab initio* Simulation Package code (see Methods)[20], as depicted in Fig. 1c, showing the valence band maximum and the conduction band minimum (CBM) located at the Γ and Λ points of BZ, respectively. The TA results measured in the 450–1400 nm region (see Methods) with excitation fluence of 90 μJ/cm$^2$ is presented in Fig. 1d, showing the photo-induced bleach peaks at 1230 nm (PIB-A), 950 nm (PIB-B), 736 nm (PIB-A′), and 607 nm (PIB-B′), which rise from the exciton transitions split by spin-orbit coupling at the K and Γ points, together with the peak at 501 nm (PIB-C) from the direct bandgap transition at the Γ point[21].

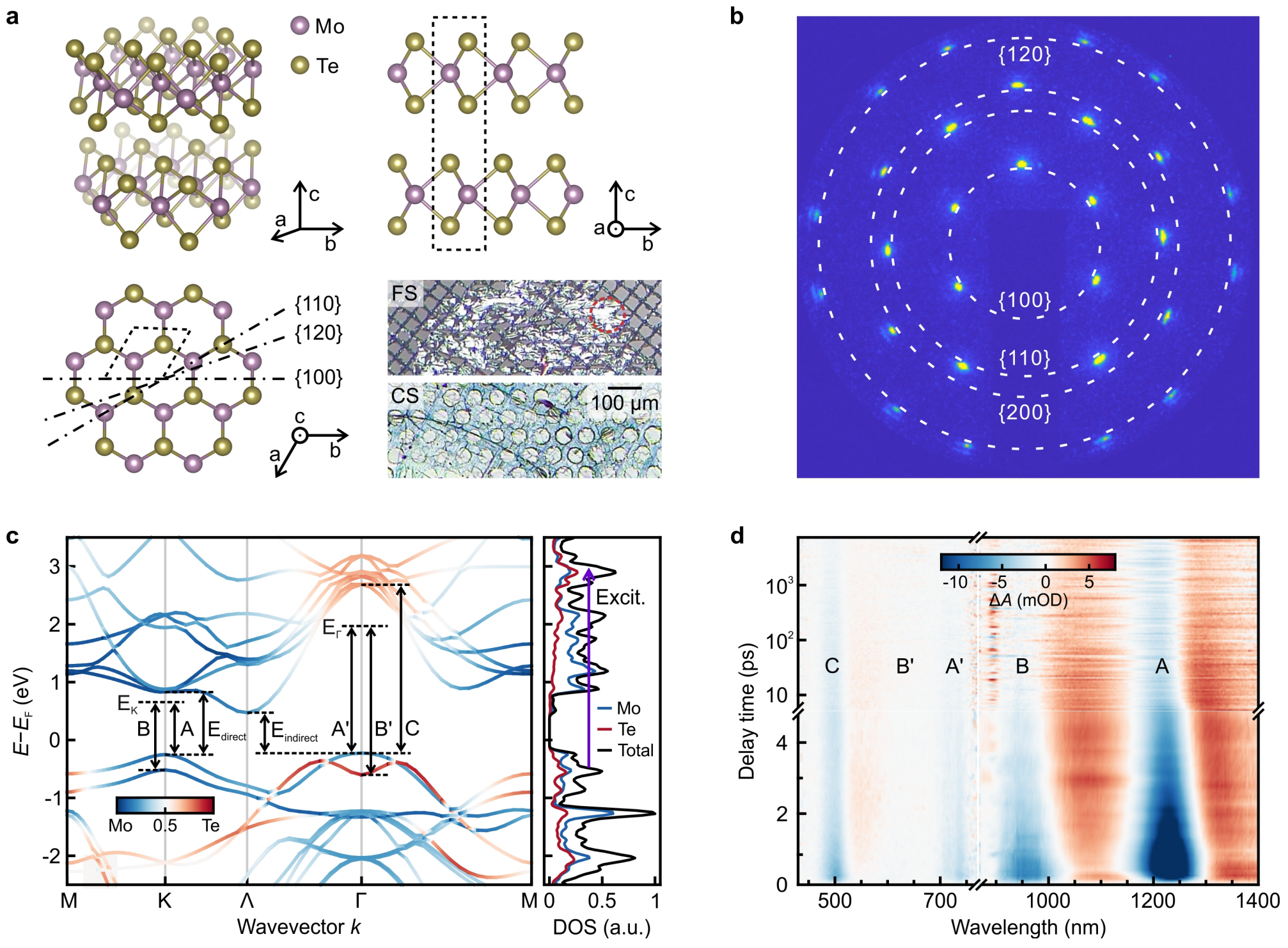

**Fig. 1 | Specimens, lattice and band structures, and UED and TA results. a**, Hexagonal lattice of 2H-$MoTe_2$ with layered structure bonding by van der Waals force, marked with the projections of crystallographic plane (dash-dot

lines) recorded in UED. Dashed frames, unit cell. Lower right panel, optical photographs of the free-standing (FS) and carbon-membrane-supported (CS) specimens, marked with the probed area (dashed red circle). **b**, UED diffraction pattern containing Bragg spots of four families of crystallographic plane parallel to the *c*-axis. **c**, Calculated electronic band structure and density of states (DOS) for bulk 2H-$MoTe_2$, marked with the transition energies under above-bandgap excitation (400 nm) probed in TA. **d**, Pseudocolor contour plot of TA spectrum in the visible–infrared wavelength range, showing photo-induced bleach peaks corresponding to the transitions marked in **c**.

## Hot carrier relaxation

We focus on the kinetics of PIB-C and PIB-A to examine the relaxations of hot carriers and band-edge excitons. The decay of PIB-C is well fitted by a tri-exponential function with characteristic times of 0.8, 5.7, and 517.6 ps, as depicted in Fig. 2a, denoting the cooling of excited electrons from the Γ point populating at the border of Λ valleys. The 0.8 ps decay is attributed to the spontaneous relaxation of hot electrons, which are described as intravalley scattering[22] towards the valley bottom, accompanied with the multiple emissions of OPs near BZ center. One SCP mode can dominate the electron-phonon scattering during intraband relaxation in TMDCs, giving rise to the phonon cascade[23] which is common in semiconductors[24], as the process I illustrated in Fig. 2b. The longer lifetime compared to the few-hundred femtoseconds reported for $MoS_2$[25] is probably slowed by the phonon bottleneck originated from phonon cascade. Based on the computed phonon dispersion curve, the highest energy of OPs at the Γ point is less than 40 meV (see Supplementary Fig. S3). Given that the energy span between the hot electrons and the Λ valley bottom is more than 1 eV, one electron needs to emit more than 25 phonons to relax to the CBM, with the emission and absorption rates of a phonon mode as[26]

$$\mathit{\Gamma}_{\mathrm{em}} \sim N_{\mathrm{P}} + 1, \tag{1}$$

$$\mathit{\Gamma}_{\mathrm{abs}} \sim N_{\mathrm{P}}, \tag{2}$$

where $N_{\mathrm{P}}$ is the population. These two rates become comparable when the population is pronouncedly larger than 1, resulting in the abundant nonequilibrium OPs emitted during phonon cascade, which in turn facilitate the reabsorption of phonons by the carriers, consequently retarding the carrier relaxation, i.e., the occurrence of phonon bottleneck.

The 5.7 ps decay is attributed to the intervalley scattering, because the six Λ valleys with surrounding satellite valleys in the CBM limit the wavevector of phonons participating in the carrier scattering, thus the phonon cascade can be interrupted only by scattering with the phonons with large enough wavevector, as the process II in Fig. 2b. The carrier cooling is possibly further slowed down effectively due to the electron storing in the satellite valleys for several picoseconds, consistent with the reported intervalley scattering of picoseconds[27] dominated by APs[28] in $MoS_2$. The last decay of 517.6 ps is attributed to the carrier recombination and diffusion, which are also reported with similar timescale in $MoS_2$[29]. This decay is represented by the indirect recombination of band-edge excitons, as the process

III in Fig. 2b, impacting the carrier relaxation through Pauli blocking[27].

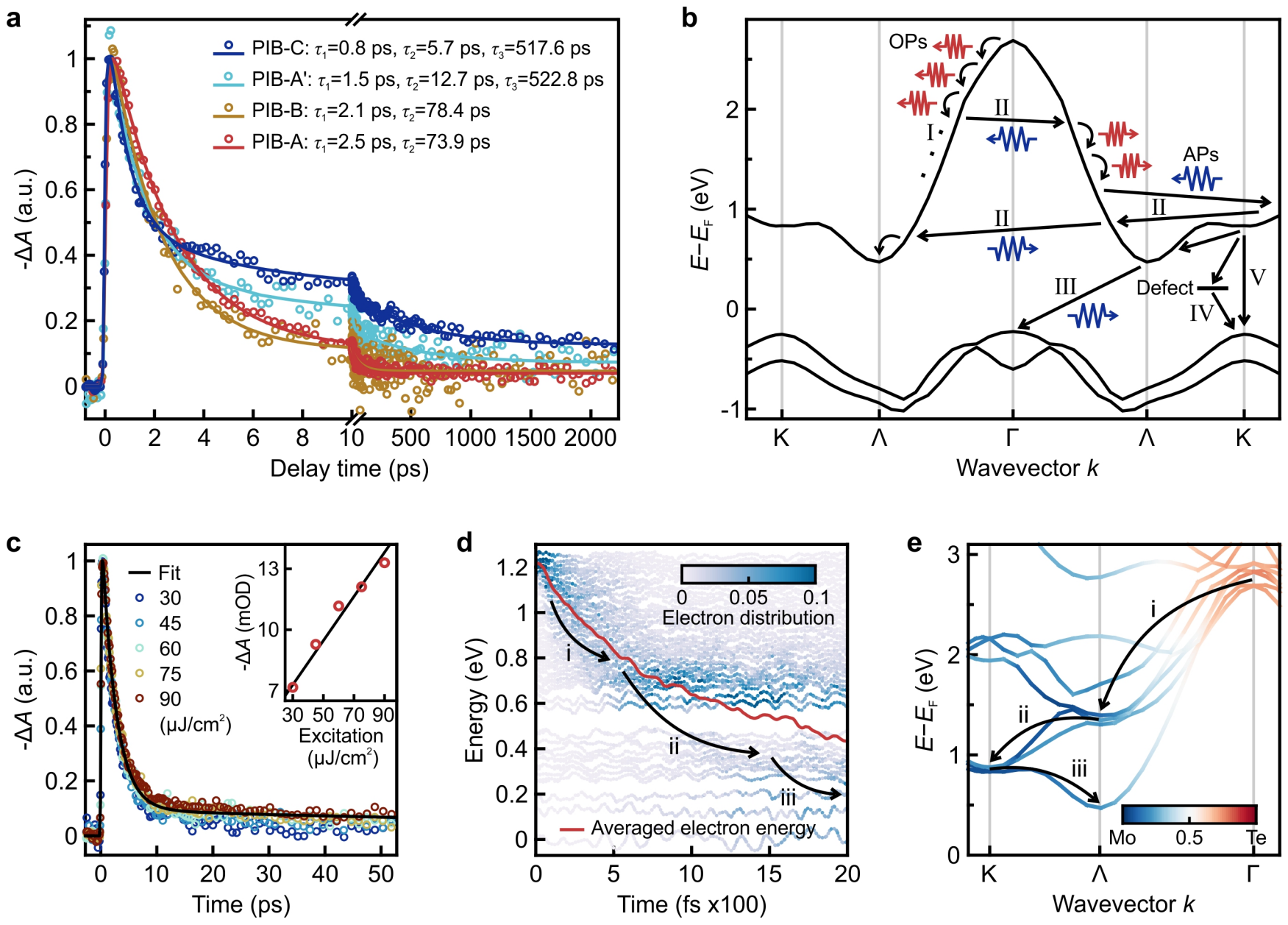

**Fig. 2 | TA measured carrier relaxations and NAMD simulations. a**, Normalized decay of photo-induced bleach peaks. Solid lines, multiexponential fits. **b**, Schematic illustration of carrier relaxation pathways in the simplified electronic bands. I, intravalley scattering accompanied with emission of optical phonons (OPs). II, intervalley scattering accompanied with emission of acoustic phonons (APs). III, indirect recombination. IV, defect-assisted recombination. V, direct recombination. **c**, Normalized decays of PIB-A under various excitation fluences, with a biexponential fit. Inset: Linear dependence of bleach amplitude on excitation fluence. **d**, Simulated three-stage energy decay of $\Gamma_e$. Processes i, ii, and iii, see Supplementary Note 1. **e**, Cooling pathways of $\Gamma_e$ in the conduction band at 0 K, concluded from the NAMD simulations. i, intravalley scattering in the Λ valley. ii and iii, intervalley scattering between the Λ and K valleys.

PIB-A decays with two processes of 2.5 and 73.9 ps. The absence of sub-picosecond process reflects no intravalley scattering for the band-edge A excitons, which populate at the bottom of K valleys. In the single particle picture, electrons at K valleys can relax to the Λ valleys though intervalley scattering. Thus the direct A excitons probably relax to indirect excitons through exciton-phonon scattering[30], forming K-Λ excitons because the bottom of Λ valley is the CBM. Such relaxations are consistent with the reduced quantum yield of photoluminescence[31]. The amplitudes of PIB-A show a linear dependence on excitation fluence, as depicted in the inset of Fig. 2c, indicating the high-order interactions such as Auger process did not play a significant role in our measurements. We therefore attributed the 2.5 ps decay to the contributions majorly from the intervalley scattering due to the low density of defect in the monocrystalline specimens, and partially from the defect-assisted nonradiative recombination[14] which is illustrated as the process IV in Fig. 2b. The 73.9 ps decay is attributed to the exciton lifetime, consistent with the reported rate of direct electron-hole recombination in other TMDCs[32], as the process V in Fig. 2b.

The concluded relaxation pathways of photoexcited carriers are corroborated by the NAMD simulations (see Methods) of the cooling and recombination of an electron with the lowest energy ($\Gamma_e$) and a hole with the second-lowest energy ($\Gamma_h$), at the $\Gamma$ points in conduction and valence bands, respectively. The simulations reveal the carrier transfers between Te and Mo atoms, and the intra- and intervalley scattering of $\Gamma_e$ and $\Gamma_h$, accompanied with the emission of $A_{1g}$ phonons. A three-stage energy decay is found for $\Gamma_e$, as depicted in Fig. 2d, confirming the observed kinetics of PIB-C. The energy decay, together with the averaged coupling along NAMD trajectory and the time-dependent spatial localization, establish the scenario of $\Gamma_e$ cooling with accurate band structure, as illustrated in Fig. 2e, well reproducing the pathways deduced from TA measurements, see Supplementary Note 1 for details.

## Coherent phonon dynamics

We employed the high sensitivity of transient absorption measured with single-wavelength (see Methods) to track the coherent phonons, which modulate the optical susceptibility thus the energy band of specimen[33]. After subtracting the exponential transition, the kinetic trace of recorded transmission shows prominent oscillations comprising of two oscillatory components, which last more than 10 ps, as depicted in Fig. 3a. The Fourier transform yields a strong peak at 5.16 THz and a weak peak at 0.19 THz, as depicted in Fig. 3b. The former is contributed by the $A_{1g}$ phonons[34] excited through strong electron-phonon coupling at $\Gamma$ point, dominating the intravalley scattering because of their vanishing momentum. The latter is contributed by the strain wave of breathing mode propagating between the two surfaces of specimen with a velocity of 2660 m/s, agreeing with the theoretically predicted 2800 m/s[35].

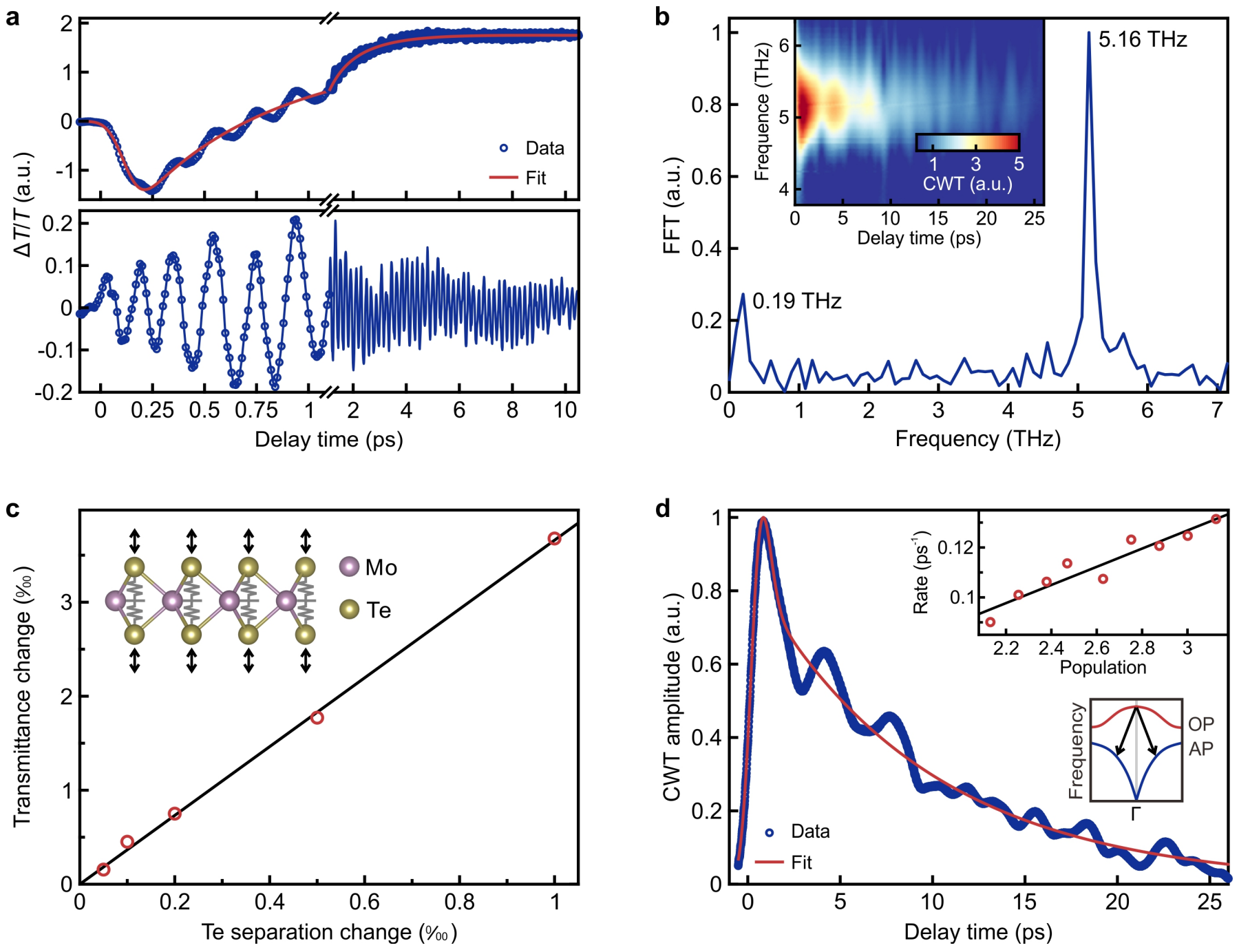

**Fig. 3 | Population and decay of coherent phonons. a**, Temporal trace of transmission change (upper panel) and the

extracted oscillation signals (lower panel), measured under 400 nm excitation with fluence of 90 μJ/cm$^2$. Only the kinetics within 10 ps are shown for clarity. **b**, Breathing mode of 0.19 THz and $A_{1g}$ mode of 5.16 THz extracted from **a** by fast Fourier transform (FFT). Inset: Continuous wavelet transform (CWT) of the $A_{1g}$ mode in **a**. **c**, Simulated linear dependence of transmittance changes on the separation changes of Te pairs, taking into account that the Te atoms vibrate along the out-of-plane direction while the Mo atoms keep still in $A_{1g}$ mode, as illustrated in the inset. **d**, Two-stage decay of $A_{1g}$ mode extracted from the inset of **b**, with biexponential fit. Upper inset: Linear dependence of extracted anharmonic rate on the calculated population of low-frequency APs, which are generated through OPs decaying, i.e., the Klemens channel illustrated in the lower inset.

Given that only the fully symmetric mode is excited and 2H-$MoTe_2$ absorbs well the laser pulses of 400 nm, the displacive excitation is probably the dominant mechanism for coherent phonon generation, as in other TMDCs[36], in which the oscillation amplitude reaches the maximum in the first period. The observed $A_{1g}$ signal in Fig. 3a takes at least 4 periods for reaching the maximum, suggesting new emissions of coherent $A_{1g}$ during the carrier relaxation. We speculated that the intravalley scattering maintains in-phase during the phonon cascade, leading to the amplification of $A_{1g}$ oscillation. Such a picture is supported by the reported more periods needed for coherent phonon reaching the maximum amplitude at lower temperature, due to the suppression of dephasing effect induced by thermal motions[37]. The measured 776 fs of 4 periods is very close to the 0.8 ps of intravalley scattering in the PIB-C decay, setting up the lower bound of time span for the dephasing surpassing the coherent emissions. At this point equations (1) and (2) yield a phonon absorption rate 0.8 times of the emission rate, giving rise to the phonon bottleneck. Considering the carrier density injected by excitation, the population of $A_{1g}$ is far more than 4. We simulated the transmittance changes induced by $A_{1g}$ vibration through calculating the relative permittivity changes with the transient structures[36], finding the transmittance changes are linearly proportional to the separation changes of Te atom pairs, as depicted in Fig. 3c. The population of $A_{1g}$ is in turn calculated using the harmonic oscillator model, yielding a value of ~$1.87\times10^7$ (see Supplementary Note 2). The estimated $N_p$ far more than 1 suggests a substantial optical phonon bottleneck, which may facilitate the high-order coupling within carriers, e.g., carrier multiplication[38].

A long lifetime of $A_{1g}$ is obtained by the analysis of continuous wavelet transform (CWT), as shown in the inset of Fig. 3b. We extracted the amplitude of CWT, finding that $A_{1g}$ decays through a two-stage process, as depicted in Fig. 3d. The dephasing of coherent phonons includes the processes of pure dephasing and population decay[39], involving changes of only momenta in the former (elastic scattering), and of both momenta and energies in the latter (anharmonic processes, e.g., three-phonon interactions). Thus, we assigned the fast decay (less than 0.3 ps) to the pure dephasing, while the slow one with characteristic time of ~10 ps to the population decay. With the fluence dependent measurements and the inclusion of common Klemens channel[40], a linear dependence of anharmonic rate on the population of lower-frequency acoustic phonons is obtained, as depicted in the upper inset of Fig. 3d. Such a linearity testifies the anharmonicity in the slow decay process, see Supplementary Note 2 for details. The slow relaxation of $A_{1g}$ facilitates the phonon reabsorption by carriers, again corroborating the picture of phonon bottleneck.

**Phonon dynamics from Brillouin zone center to border**

More phonon dynamics are encoded in the lattice responses captured by UED (see Methods), as the phonon populations alter the scattering of probe electrons. The intensity kinetics of recorded Bragg spots of the FS and CS specimens are depicted in Fig. 4a, showing decay processes with four and three stages, respectively. The FS specimen first undergoes two sequential decays with characteristic times of ~2 ps and ~30 ps, respectively, in which the relative intensity changes follow the Debye-Waller description[41], as depicted in Fig. 4b, indicating the lattice thermalization. During the next ~100 ps, the FS specimen reaches a quasi-thermal-equilibrium, demonstrating the process of phonon-phonon scattering among different modes. Note that the phonons captured in UED exclude the $A_{1g}$ mode, which vibrates parallel to the propagation direction of probe electron beam. A third intensity decay of ~900 ps follows the quasi-equilibrium, suggesting an extra thermalization process. We attributed this long decay to the nonradiative recombination of band-edge excitons, which was also detected in the carrier relaxation with similar timescale depicted in Fig. 2a. At last, the lattice begins to cool down after 2 ns of heating up, indicated by the intensity recovery in ~4 ns (limited by the time window of measurement). The CS specimen undergoes two similar intensity decays of ~2 ps and ~20 ps, followed by a lattice cooling of ~700 ps. The accelerated thermalizing and cooling processes demonstrate the phonon coupling across the heterostructure interface, see Supplementary Note 3 and 4 for further discussions.

Addition to the accelerated rates, the amplitude of first decay in the CS specimen (8.3%) is apparently less than that in the FS specimen (13.1%), indicating the decreased population of SCPs due to the extra dissipation channels provided by the substrate. We thus conclude that the acoustic SCPs, besides the $A_{1g}$, also populate and reach a non-equilibrium distribution in the excited 2H-$MoTe_2$, because the group velocity of OPs in the 2H-$MoTe_2$ film is too slow to directly couple to the phonons in substrate[26]. Taking the quasi-ballistic effect and the thermal boundary resistance into account, we deduced a thermal boundary conductance of ~10 $MW \cdot m^{-2}K^{-1}$, which is close to the value of $MoS_2$ film with $SiO_2$ substrate[42], well modeling the experimental conditions, see Supplementary Note 4 for details.

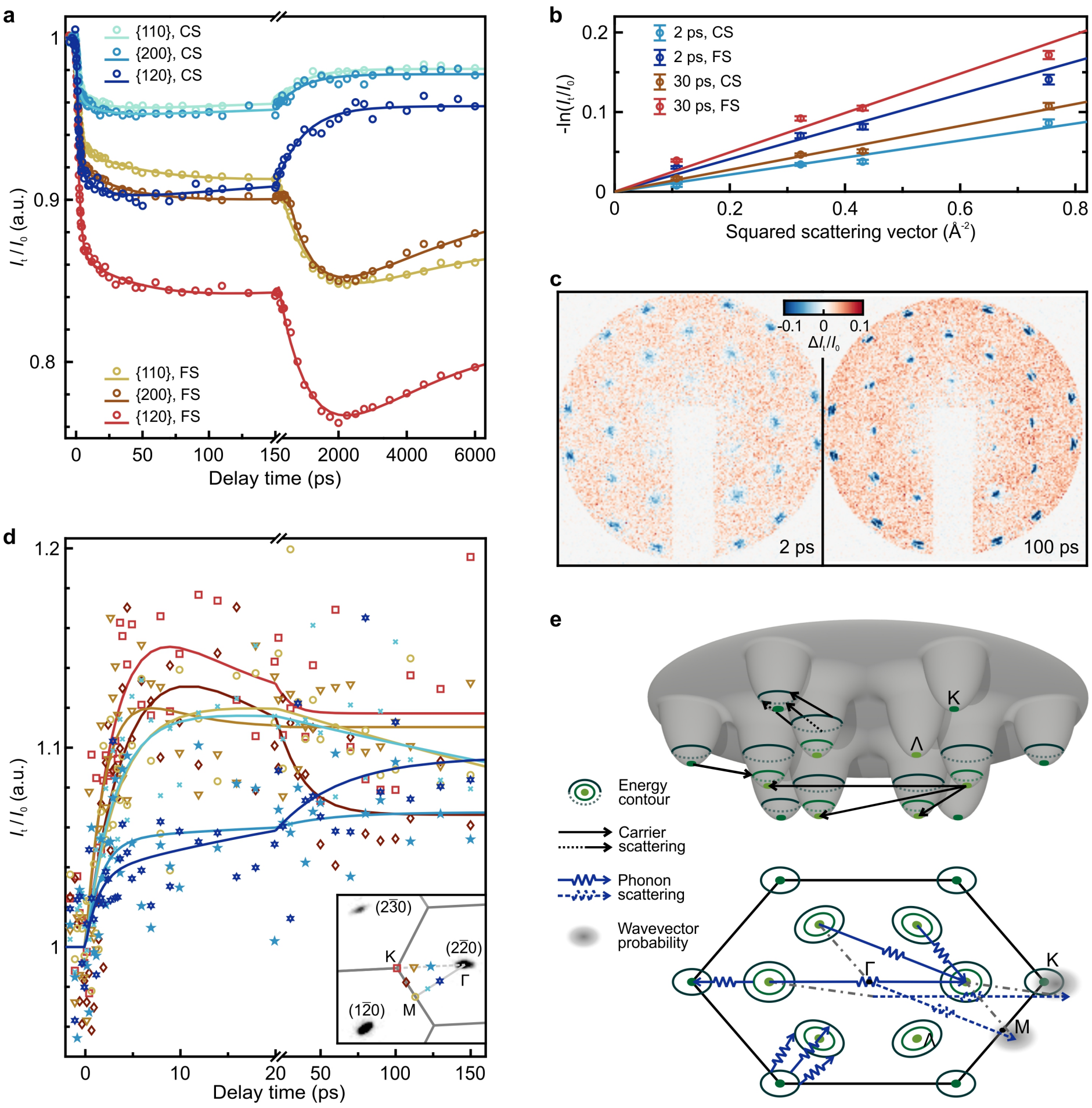

**Fig. 4 | Lattice dynamics and intervalley scattering of carriers and phonons. a**, Intensity decay of Bragg spots under excitation of 120 μJ/cm$^2$ in the FS and CS specimens. **b**, Linear dependence of relative intensity changes on squared scattering vector for the first two decays in **a**. **c**, Differential diffraction patterns of FS specimen at delay times of 2 (left) and 100 (right) ps, referred to the pattern before excitation. Both the increase (red) of diffuse scattering and the decrease (blue) of Bragg scattering indicate the populating phonons. **d**, Kinetics of diffuse signal at the selected locations marked in the inset. Note that the low signal-to-noise rate is due to the very low intensity of diffuse scattering (a factor of $10^{-3}$ compared to the signal of Bragg spots). Solid lines, multiexponential fits. **e**, Schematic illustration of intervalley scattering of carriers and phonons in the reduced and projected band structures with six Λ and six K valleys located with rotational symmetry. Dash-dot lines, the guides to the eye for equivalent transitions.

Acoustic SCPs at the M and K points of BZ dominating the ultrafast lattice disorder are reported in other TMDCs[18,43]. In the description of electron-phonon coupling with structured susceptibility, SCPs arise from the wavevector-dependent coupling vertex[44], because they correspond to the inelastic scattering and the disappearance of Fermi surface nesting in bulk 2H-$MoTe_2$, see Supplementary Note 5 for further discussions. The larger coupling constant of longitudinal APs at the M point (LAPs-M) compared to those of other acoustic modes is revealed by the simulation for 2H-$MoTe_2$[11]. So that we believe the first intensity decay is majorly contributed from the population of LAPs-M, which follows the Debye-Waller relationship[45]. The populations in FS and CS specimens are estimated to be ~$8\times10^{12}$ and ~$5\times10^{12}$, respectively (see Supplementary Note 6), suggesting the occurrence of phonon bottleneck of LAPs-M.

Compared to the Bragg scattering encoding the momenta transferred from reciprocal vectors, the signals of diffuse scattering encode the momenta transferred from phonons at specific BZ locations[46]. The differential patterns of FS specimen at 2 and 100 ps referred to that before excitation are shown in Fig. 4c, showing diffuse signals in the former barely changed near the reduced Bragg spots (centers of BZ) while increased at other locations. In the latter, the diffuse signals uniformly increased, with further reduction of Bragg spots. The delayed diffuse signal increase at the locations near Bragg spot suggests the stronger electron-phonon coupling at BZ border compared to that near the BZ center, inducing the faster phonon population at locations away from the Γ point. The diffuse signal kinetics at selected locations are depicted in Fig. 4d, representing the phonon dynamics extracted from the high-symmetry points and the locations close to BZ center, as marked in the inset of Fig. 4d. All kinetics show identical fast intensity increases (~3 ps) followed by distinct second processes, which exhibit a trend of intensity change from decrease to increase as the location changing from the BZ border to the center. Such evolutions demonstrate the energy relaxation pathway from hot carriers to SCPs, then to low-frequency phonons.

In TMDCs, interlayer interactions result in the Davydov splitting[47] of acoustic phonon modes, forming the ultralow-frequency region in the phonon spectra. We examined the scattering of LAPs-M in this region. The phonons at locations around M points, with the eigenvectors shown in Supplementary Fig. S9, are expected to possess high scattering rates due to the spin texture[48,49]. We estimated the nonequilibrium population of LAPs-M by implementing the variation of first-order diffuse intensity, and excluding the contributions from high frequency phonons and higher-order scattering for their much smaller cross sections. The calculations yielded a lower bound of 54 (see Supplementary Note 7), again suggesting the phonon bottleneck of LAPs-M.

A more detailed diagram of intervalley scattering of carriers and phonons is schematically illustrated in Fig. 4e, showing the phonon emissions around K points induced by the electron scattering between opposite Λ valleys, and those around M points by the electron scattering between straddling Λ valleys. These phonons are equivalent to those emitted from the Γ point to the BZ border, contributing to the diffuse signals at high-symmetry points. All these intervalley scatterings are involved in the APs emission illustrated in Fig. 2b, dominating the fast intensity decay of Bragg spots.

## Discussion

Combining the results of calculation and measurements of TA and UED, we are able to construct a comprehensive picture of energy relaxation after photoexcitation in 2H-$MoTe_2$, as illustrated in Fig. 5.

The photoexcited hot carriers decay through intravalley scattering within a subpicosecond duration, accompanied with the emissions of $A_{1g}$ OPs. A part of the hot carriers decay through intervalley scattering within several picoseconds, accompanied with the APs emission preferentially at the BZ border, due to the obstruction in intravalley cooling induced by the phonon bottleneck of $A_{1g}$. The acoustic SCPs also give rise to the phonon bottleneck at BZ border within a few picoseconds, which is followed by the population of low-frequency phonons in couple tens of picosecond, resulting in the fast disorder of lattice. The FS specimen maintains a quasi-thermal-equilibrium lasting a hundred picoseconds, while the CS specimen cools down several nanoseconds earlier because the substrate effectively drains away the acoustic phonons prior to their equilibrium, therefore alleviates the obstruction of thermal relaxation.

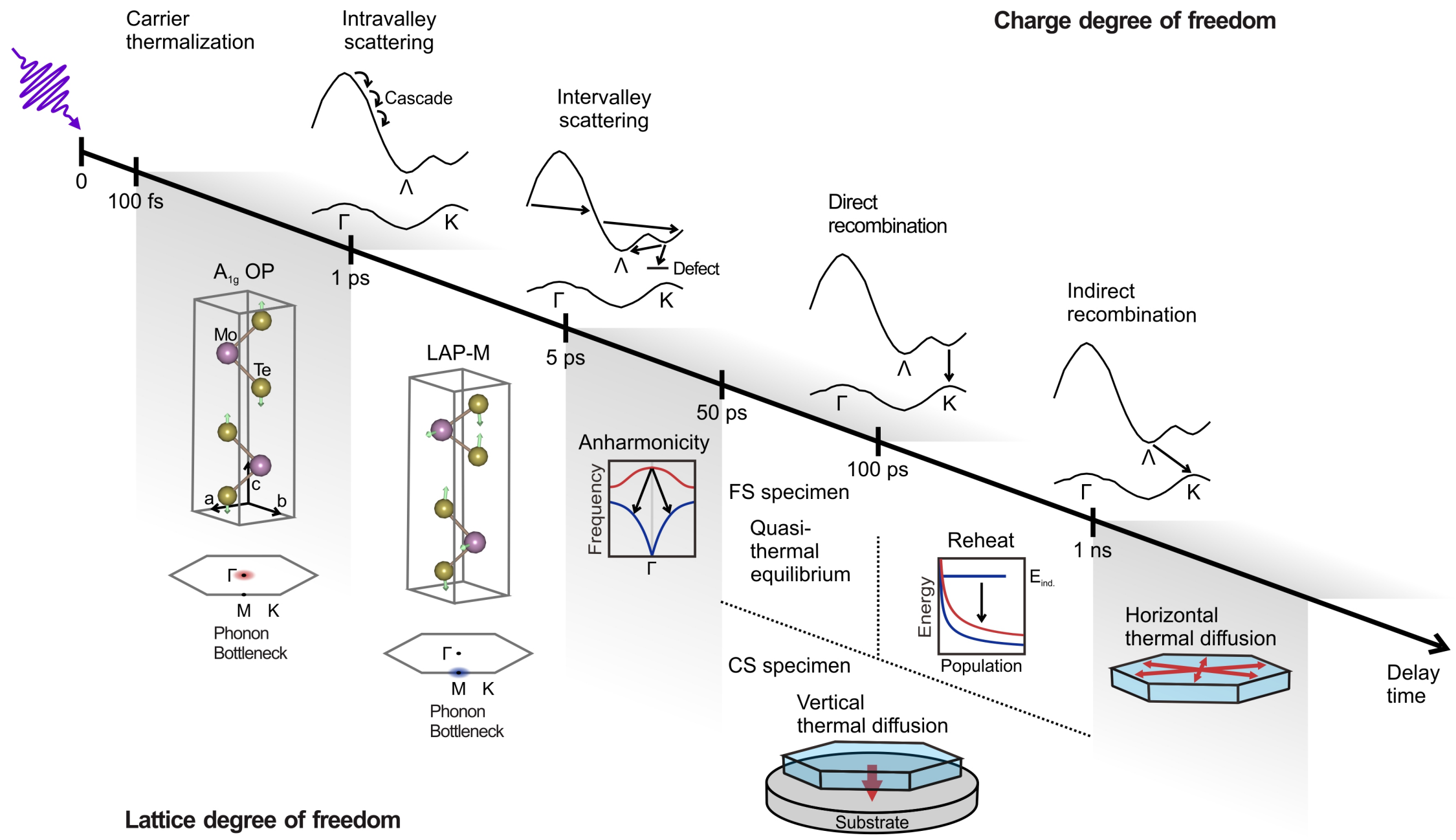

**Fig. 5 | Schematic diagram of photoexcited carriers and phonons evolving in 2H-$MoTe_2$ during a time window of nanoseconds.**

The strong electron-phonon coupling in 2H-$MoTe_2$ is unprofitable for applications utilizing excited carriers, but the phonon bottlenecks and the carrier multiplication provide potentials for the application of hot-carrier solar cell with improving photoelectric conversion efficiency[50]. As either photon or electric field stimulates phonon populations in semiconductors[26], the rich phonon pathways revealed in 2H-$MoTe_2$ also provide potentials for heat management applications.

The implement of ultrafast optical and electron probes, with the advantage of momentum-resolving in the latter, makes possible picking out the information of phonon bottleneck from the shielding effect[51] of electron-phonon coupling, and from the different valleys in band structure. We demonstrate here various approaches to estimate the phonon populations based on the measurements of optical absorption and electron diffraction, quantitatively distinguishing the occurrence of different phonon bottlenecks. At the end, we are able to elucidate the carrier and phonon interactions across the charge, lattice, and valley degrees of freedom on their respective timescales, providing a frame for better understanding this highly complex scenario, and hopefully for harnessing these processes in valleytronics and other optoelectronic applications.

## Methods

**Specimen preparation.** The Mo film deposited on a 1-inch Si/$SiO_2$ wafer through magnetron sputtering was tellurized with Te vapor in a horizontal hot-wall tube furnace (at ~510 ºC) to obtain a polycrystalline 1T′-$MoTe_2$ film. A seed single-crystalline 2H-$MoTe_2$ nano-flake, mechanically exfoliated from the bulk crystal, was dryly transferred on the 1T′-$MoTe_2$ film. After that, an $Al_2O_3$ layer was deposited on the surface by atomic layer deposition, and a small hole through 2H- and 1T′-$MoTe_2$ was punched by a tungsten probe mounted on a 3D translation stage. The wafer and a 3-g Te lump were placed in a closed quartz tube and then loaded into another quartz tube in a chemical vapor deposition furnace. The 1T′-$MoTe_2$ layer underneath the seed crystal transformed into a 2H-$MoTe_2$ single crystal through the phase transition and recrystallization induced by interface at ~650 ºC. At last, the $Al_2O_3$ layer was dissolved using a hot phosphoric acid solution. After etching the Si/$SiO_2$ substrate by hydrogen fluoride, the 2H-$MoTe_2$ film can float on liquid level. Then the copper grids, bare or coated with ultrathin amorphous carbon membrane (with mean thickness of ~8.5 nm), were used to attach the 2H-$MoTe_2$ films as FS and CS specimens, as shown in Fig. 1a. More details can be found in the earlier report[19]. The atom-resolved structure and crystallization of the obtained specimens are characterized by the electron imaging and diffraction, as shown in Supplementary Fig. S1.

**Density functional theory (DFT) calculation.** The calculations of geometry optimization, electronic band structure, optical properties, and Fermi surface were performed using DFT within the Vienna *Ab initio* Simulation Package (VASP) code[52] with the generalized gradient approximation of Perdew–Burke–Ernzerhof (GGA-PBE). The projected augmented wave method was utilized with a plane-wave basis set[53,54], with the convergence criteria set to $10^{-5}$ eV for energy and 0.01 eV/Å for force. A kinetic cutoff energy of 500 eV and a Monkhorst-Pack[55] special k-point mesh of 5×5×1 were employed for calculations. Additionally, the DFT-D3 method developed by Grimme was employed to account for the van der Waals interactions[55]. The vibrational properties are obtained from the density functional perturbation theory using the Phonopy code[56] and a 4×4 supercell.

**Non-Adiabatic Molecular Dynamics (NAMD)**. The Hefei-NAMD code[57,58] was employed for the NAMD simulation. The relaxation of photoexcited carriers was modeled using the decoherence-induced surface hopping method, within the framework of time-dependent density functional theory in the Kohn-Sham (KS) scheme[59]. Prior to conduct the NAMD calculations, the adiabatic molecular dynamics (MD) simulations were performed with an orthogonal 3×3 supercell, consisting of a total of 108 atoms. The structure was fully relaxed at 0 K, then heated to 300 K over a period of 1 ps using repeated velocity rescaling. Following the heating phase, an adiabatic MD trajectory of 5 ps was generated in the microcanonical ensemble (NVE) with a nuclear time step of 1 fs. The nonadiabatic coupling matrix elements were computed along the trajectory. To simulate the electron/hole dynamics, 200 initial configurations were selected from the final 3 ps of the trajectory.

**Transient absorption spectroscopy.** The TA measurements were performed under the ambient atmosphere using a Helios spectrometer (Ultrafast System), and a Ti:sapphire regenerative amplifier (Legend, Coherent) which delivers pulses of 800 nm, with pulse width of 40 fs and repetition rate of 5 kHz. The repetition rate was chopped to 2.5 kHz in measurements. The output from the regenerative amplifier was divided into two arms by a beam splitter. One arm was used for the second harmonic generation from a $BaB_2O_4$ (BBO) crystal to excite the specimen, so that the 3.1 eV photons drove the population of hot carriers and the subsequent carrier multiplication. The other arm was used for the generation of white-light continuum (Ti:sapphire for 400–800 nm, Nd:YAG for 850–1450 nm) as the probe pulses. The diameter of pump and probe spots were 200 μm and 100 μm, respectively. The instrument response function was measured as ~100 fs.

**Single-wavelength transient absorption measurement.** The specimen was excited with the same conditions as the TA measurement, and probed by the fundamental output of 800 nm from the Ti:sapphire regenerative amplifier. The repetition rate was chopped to 125 Hz in measurements. A silicon detector (DET36A/M) was used to record the transmitted light, then the signals were amplified by a lock-in amplifier (SR865, Stanford Research Systems).

**Ultrafast electron diffraction.** The UED measurements were performed under the ultrahigh vacuum condition using a home-built ultrafast electron diffractometer with estimated temporal resolution of subpicosecond[60], and the same Ti:sapphire regenerative amplifier as the TA measurement. The specimen was also excited with 400 nm pulses. The pulsed probe electrons were accelerated by an electric field of 30 kV, generating diffraction patterns through a transmission geometric setup. The diffraction signals were gained by two chevron-stack microchannel plates, then recorded by a COMS camera (ORCA-Flash, Hamamatsu). Each pattern was recorded with 15000 accumulated pulses. The diameter of pump and probe spots were ~700 μm and ~100 μm, respectively. The intensity of Bragg spots presented in the manuscript is averaged over all recorded spots of the same family.

**Acknowledgements**

Z.W. and C.H. thank the financial support from the National Natural Science Foundation of China

(62474073). We thank Dr. Z.X. for the helps on DFT calculations.

## Author contributions

W.L. conceived of and supervised the project. Z.W. performed the measurements with supports from Y.S., X.W., Z.Z., X.Z., F.H., Q.Y. and C.H. Y.P synthesized the 2H-$MoTe_2$ films under supervisions from Y.Y. M.L. performed the NAMD simulation under supervisions from J.H. Z.W. and W.L. analyzed the data with discussions with all authors. Z.W. and W.L. wrote the manuscript with contributions from all authors.

## Competing interests

The authors declare no competing interests.

## References

### main

## Methods

Supplementary Information

# Valley resolved dynamics of phonon bottleneck in semiconductor molybdenum ditelluride

Zhong Wang[1,2], Yijie Shi[1,2], Yu Pan[3], Min Li[4], Xi Wang[1,2], Zheng Zhang[1,2], Xiangyu Zhu[1,2], Fuyong Hua[1,2], Qian You[1,2,†], Chunlong Hu[1, 2], Junjie He[4], Yu Ye[3], Wenxi Liang[1,2,*]

[1]Wuhan National Laboratory for Optoelectronics, Huazhong University of Science and Technology, 1037 Luoyu Road, Wuhan 430074, China

[2]Advanced Biomedical Imaging Facility, Huazhong University of Science and Technology, Wuhan 430074, China

[3]State Key Laboratory for Mesoscopic Physics and Frontiers Science Center for Nano-Optoelectronics, School of Physics, Peking University, Beijing 100871, China

[4]Department of Physical and Macromolecular Chemistry, Faculty of Science, Charles University, Prague 12843, Czech Republic

[†]Current address: Zhongshan Lighting Fast Intellectual Property Rights Service Center, Zhongshan 528421, China

[*]Email: wxliang@hust.edu.cn

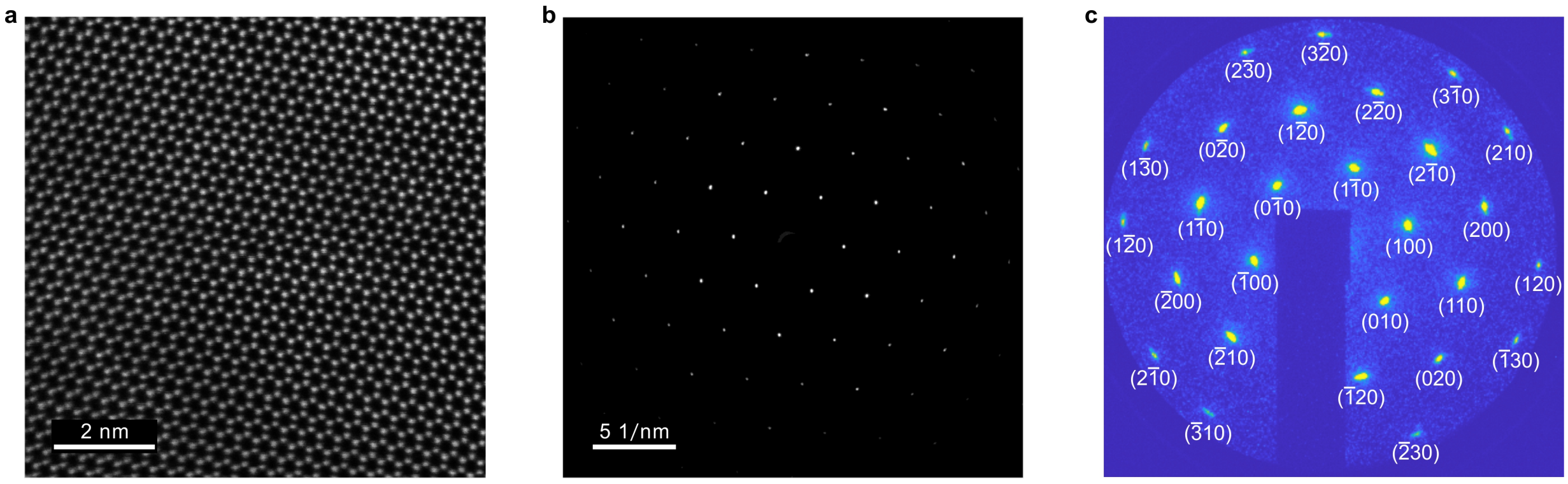

**Supplementary Figure S1: Characterizations of Specimen. a**, Scanning transmission electron microscope (JEM-ARM200F, JEOL) image of the free-standing (FS) specimen, showing regular hexagon grids composed of Mo and Te atoms with no obvious contrast of different atoms, due to the similar atomic numbers and the overlap of different atoms in different layers. **b**, Selected area electron diffraction image, with sharp Bragg spots manifesting the high quality of crystallization. **c**, Electron diffraction pattern of the carbon-membrane-supported (CS) specimen recorded by the homebuilt ultrafast electron diffractometer (UED), with corresponding Miller indices.

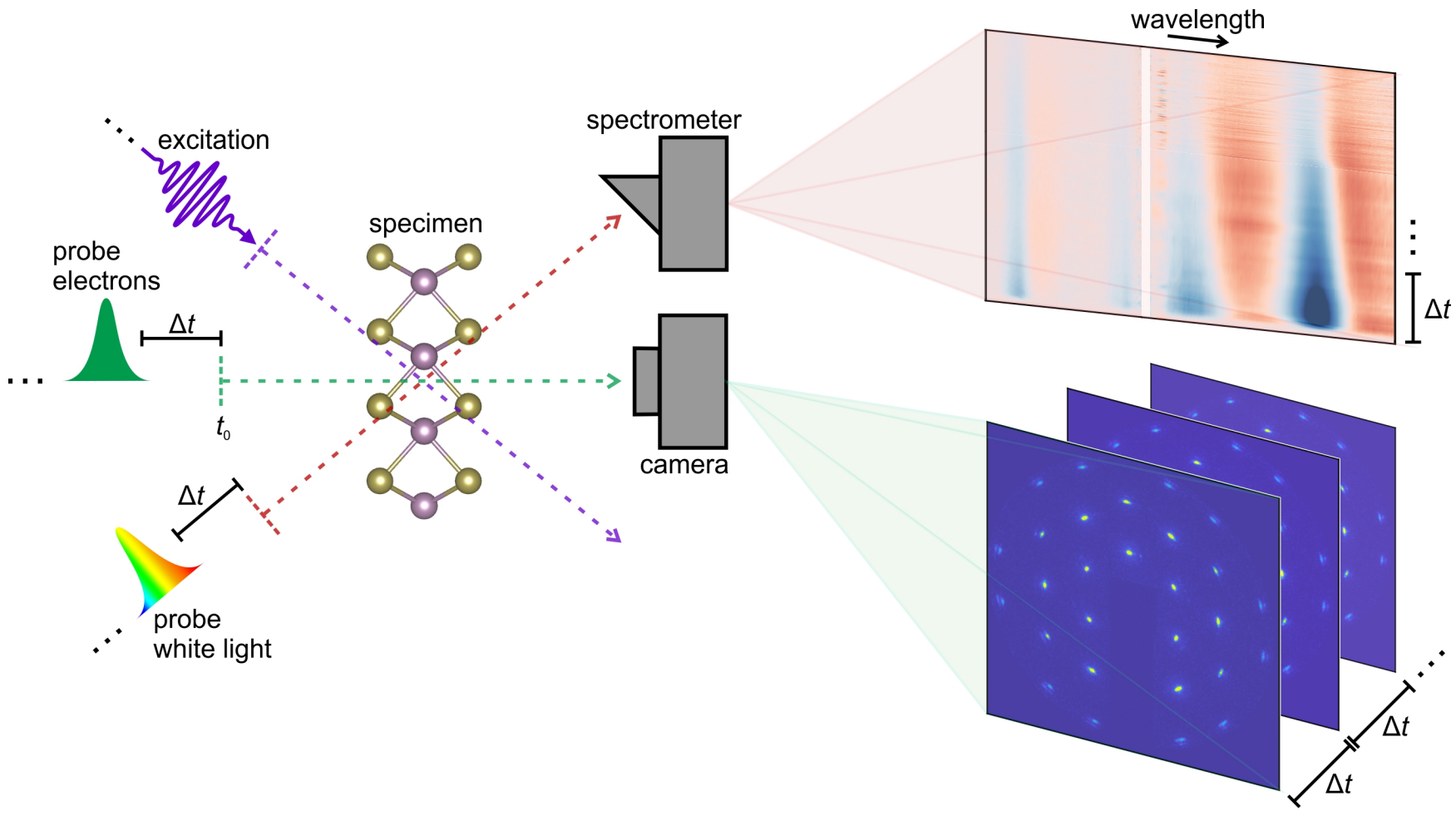

**Supplementary Figure S2: Schematic diagram of experimental setup**. The pump-probe based transient absorption (TA) spectroscopy and UED record the differential absorption spectra and the diffraction patterns, respectively, at different delay times after the excitation by femtosecond laser pulses.

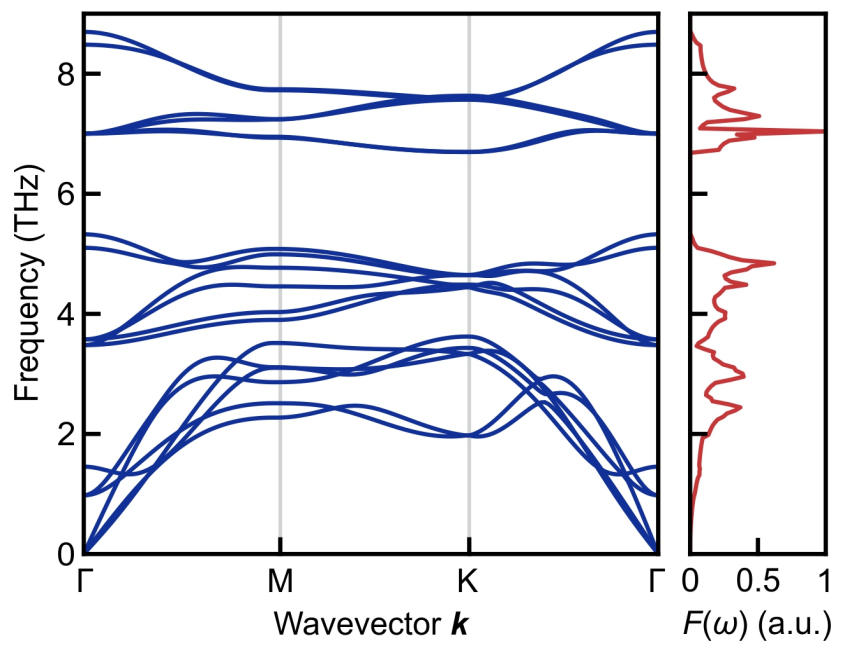

**Supplementary Figure S3: Calculated phonon dispersion curves and normalized phonon density of states $F(\omega)$.** The highest energy of optical phonons at the Γ point is ~8.7 THz (36 meV).

## Supplementary Note 1：Simulations of Non-Adiabatic Molecular Dynamics (NAMD) and the relaxation pathways of $\Gamma_e$ and $\Gamma_h$

During the relaxation of photoexcited carriers, the extent of charge transfer in 2H-$MoTe_2$ is determined by integrating the carrier density, which is expressed as[1]:

$$\int \rho(r,t)dr = \int |\Psi(r,t)|^2 dr = \sum_{i,j} c_i^*(t)c_j(t) \int \varphi_i^*[r,R(t)]\varphi_j[r,R(t)]dr\,, \qquad \text{(S1)}$$

where $\rho$ is the density of photoexcited charges, $\Psi$ is the total wave function, which can be expanded into the Kohn-Sham (KS) wave functions $\varphi_i$ and $\varphi_j$ with coefficients $c_i$ and $c_j$ indicating the occupation of excited carriers on KS orbitals, respectively. The time derivative of equation (S1) provides the expression of adiabatic (AD) and nonadiabatic (NA) contributions to charge transfer:

$$\frac{d\int \rho(r,t)dr}{dt} = \sum_{i,j}[\frac{d(c_i^* c_j)}{dt}\int \varphi_i^* \varphi_j dr + c_i^* c_j \frac{d(\varphi_i^* \varphi_j)dr}{dt}]. \quad (S2)$$

The first term on the right-hand side describes the change in charge density derived from the variation of occupation of the adiabatic KS states, referred to as the NA term. The second term accounts for the effects arising from the changes in localization of adiabatic KS states, referred to as the AD term. Integrating these two terms yields their contributions to the total charge transfer. Both terms can be promoted by increasing temperature due to the crucial role of phonon in these terms.

Using the decoherence-induced surface hopping (DISH) method[2], we simulated the cooling and recombination of $\Gamma_e$ and $\Gamma_h$, which are associated to the excitation transition (illustrated in Fig. 1c in the main text) generating the PIB-C signal (shown in Fig. 1d in the main text). The oscillations of Kohn-Sham eigenenergy of the NAMD simulation supercell, induced by the strong electron-phonon coupling at 300K, are depicted in Fig. S4a. The fast Fourier transform (FFT) of the temporal evolution of electronic states near the conduction band minimum (CBM) yields a peak of 5.3 THz, as depicted in Fig. S4b. This oscillation is attributed to the $A_{1g}$ mode[3], which couples to electrons stronger than other modes and dominates the phonon cascade illustrated in Fig. 2b in the main text.

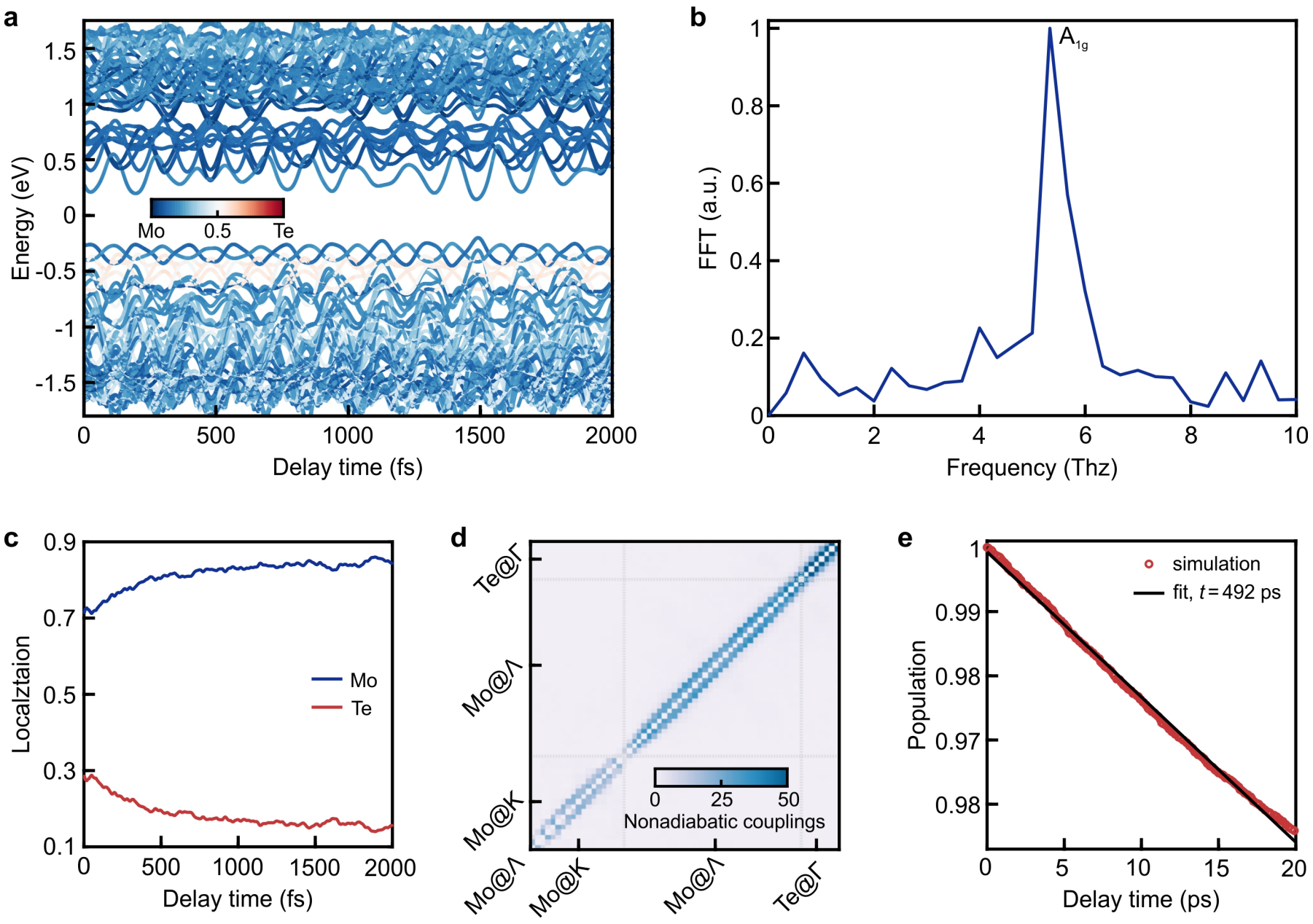

**Supplementary Figure S4: Simulations of $\Gamma_e$ relaxation. a**, Temporal evolution of energy states near the Fermi level at 300 K. The color map denotes the orbital localization at different atoms. **b**, FFT of states near the CBM. **c**, Time-dependent localization at different atoms. **d**, Averaged nonadiabatic couplings along the NAMD trajectory of 2 ps. **e**, Population decay in the conduction band.

$\Gamma_e$ transfers from Te to Mo during the first ~500 fs of cooling, staying mainly in Mo, as depicted in

Fig. S4c. The energy decay of $\Gamma_e$ consists of three continuous stages marked as i, ii, and iii, , as depicted in Fig. 2d in the main text: i, changing from the high-energy region (HER) centering at ~1.2 eV to the medium-energy region (MER) centering at ~0.8 eV within the first ~500 fs; ii, staying in the MER then changing to the low-energy region (LER) below 0.4 eV within ~500 fs to ~1.5 ps; iii, changing in the LER and further relaxing to the CBM after ~1.5 ps.

Then we considered the nonadiabatic couplings (NACs)[1] between different states, as depicted in Fig. S4d. The couplings between the Te orbitals in HER at the Γ point are strongest, resulting in the rapid transfer of $\Gamma_e$ between the Te orbitals before the i stage, which is consistent with the barely changed electron localization within the first 50 fs (see Fig. S4c). Nest, the couplings between the Te and Mo orbitals in HER at the Γ point and in MER at the Λ point are also strong, resulting in the rapid transfer of $\Gamma_e$ to the Mo orbitals during the i stage. The couplings between the Mo orbitals in MER at the Λ point and in LER at the K point are weakest, resulting in $\Gamma_e$ staying in the Mo orbitals at Λ point during the ii stage. At last, the couplings between the Mo orbitals in LER at the K and Λ points are also weak, resulting in the iii stage.

Taking all calculation and simulation results into account, we concluded the cooling pathway of $\Gamma_e$ in the actual band structure, as depicted in Fig. 2e in the main text. The i stage is attributed to the intravalley scattering in the Λ valley, which is faster than the results of TA measurement due to ignoring the complex effects, such as phonon bottleneck. The ii and iii stages are attributed to the intervalley scattering between the Λ and K valleys, in which $\Gamma_e$ takes several picoseconds, a duration close to the measured characteristic time of second decay of PIB-C, to reach the CBM. The population decay of excited carrier is depicted in Fig. S4e, in which we estimated the lifetime of electron-hole recombination as ~492 ps, close to the measured characteristic time of third decay of PIB-C, via a short-time linear approximation[4]. The simulation results testified the assignment of PIB-C to the indirect recombination.

The cooling process of $\Gamma_h$ is presented in Fig. S5. The energy decay of $\Gamma_h$ to the VBM consists of two stages marked as j and jj, as depicted in Fig. S5a: j, rapid hole transfer to the lower energy levels within ~300 fs, driven by the strong coupling between the Mo and Te orbitals at Γ point (see Fig. S5b) and accompanied with the change of localization at different atoms (see Fig. S5c), indicating $\Gamma_h$ relaxing to the energy level around Γ point; jj, unchanging averaged energy of $\Gamma_h$, suggesting the back and forth scattering between the electron energy peaks at Γ and K points, in which the energy states are almost equal. We assigned the jj stage to the intervalley scattering process, with a characteristic time of ~700 fs estimated from the decay of NA term contributing to the hole transfer, as depicted in Fig. S5c. In the end, the relaxation pathway of $\Gamma_h$ is summarized in a simplified band structure, as depicted in Fig. S5d, showing a decay process with two stages, j, the intravalley scattering within Γ valley, and jj, the intervalley scattering between the Γ and K valleys.

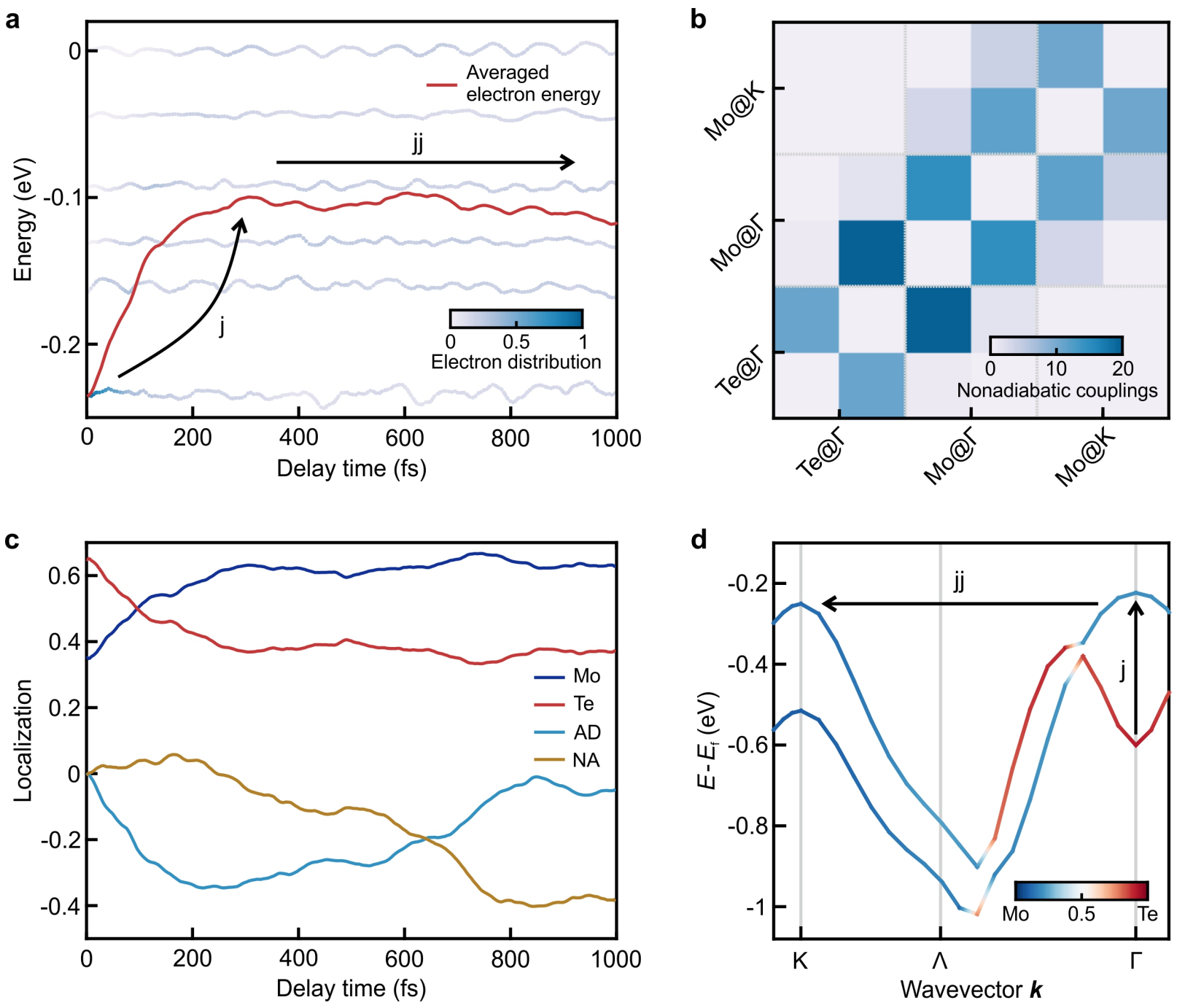

**Supplementary Figure S5: Simulations of $\Gamma_h$ relaxation. a**, Temporal evolution of energy states near the Fermi level at 300 K. The color map denotes the orbital localization at different atoms. **b**, Averaged nonadiabatic couplings along the NAMD trajectory of 1 ps. **c**, Time-dependent localization at different atoms, and AD and NA contributions to the hole transfer. **d**, Summarized relaxation pathway.

## Supplementary Note 2: Calculations of coherent phonon population and the following anharmonic process

The coherent oscillation with frequency of ~0.19 THz is attributed to the propagating strain wave. Hence the longitudinal sound velocity can be estimated as[5]:

$$v = \frac{2d_{\mathrm{M}}}{T}, \tag{S3}$$

where $d_{\mathrm{M}}$ is the thickness of specimen and $T$ is the oscillation period. The obtained 2660 m/s agree well with the reported 2800 m/s by simulation[6].

The other observed coherent oscillation of $A_{1g}$ phonon is assigned to the ZO mode, which is different from the strongly coupled longitudinal optical (LO) modes in polar materials such as perovskites[7], due to the negligible Fröhlich interaction[8]. Under the excitation of femtosecond laser pulses, coherent phonons are generated, in general, through the mechanisms of resonant impulsive stimulated Raman scattering (ISRS) or the displacive excitation of coherent phonons (DECP)[9]. ISRS allows for stimulating all Raman-active modes in the transparency regime, while DECP favors only stimulating fully symmetric modes (e.g., $A_{1g}$) in the absorption regime. Given the absorption of 400 nm excitation pulses and the absence of other Raman modes in our measurements, we reasoned that DECP played the dominant role in the generation of photo-induced coherent phonons of 2H-$MoTe_2$. Regardless of whether ISRS or DECP is the origin, coherent phonons are expected to reach the maximum amplitude in the first period of vibration. From the delayed maximum amplitude (see the lower panel of Fig. 3a in the main text) we inferred that new $A_{1g}$ emissions took place coherently after the initial one, increasing the population and therefore the amplitude of observed oscillation in later periods.

We estimated the $A_{1g}$ population through the transmittance change induced by coherent vibrations, considering the specimen in probed area is uniformly excited. In the $A_{1g}$ mode, only Te atoms vibrate along the out-of-plane direction. There are ~$1.4\times10^{14}$ Te atoms in the probed volume (a spot with diameter of 100 μm containing ten layers). By treating each Te vibration as a simple harmonic oscillator, as illustrated in the inset of Fig. 3c in the main text, we have:

$$\left(n_{\mathrm{phonon}} + \frac{1}{2}\right)\hbar\omega = \frac{1}{2} m_{\mathrm{Te}}\omega^2 {d_{\mathrm{Te}}}^2 n_{\mathrm{Te}}, \qquad (S4)$$

where $m_{\mathrm{Te}}$ is the Te atomic mass, $\omega$ is the angular frequency of $A_{1g}$ mode, $d_{\mathrm{Te}}$ is the amplitude of vibration, $n_{\mathrm{Te}}$ is the total number of Te atoms in the probed volume, and $\hbar$ is the reduced Planck constant. Equation (S4) connects the population of $A_{1g}$ ($n_{\mathrm{phonon}}$) with the vibration of Te atoms. To simulate the transmittance change, the needed absorption coefficient and reflectivity are obtained via calculating the dielectric function, which is estimated by the DFT calculation. Then the transmittances of two transient structures with Te atom pairs at the maximal and minimal separations, respectively, are calculated using the VASP code. Taking the excitation of 90 μJ/cm$^2$ as an example, the calculated vibration amplitudes correspond to a change of ~0.4‰ of the totally transmittance, with a change of ~0.11‰ of the separation of Te atom pairs. The estimated maximum population is ~$1.87\times10^7$, a value far more than 1, suggesting the occurrence of phonon bottleneck, and the possible photoinduced Te segregation[10] and nonthermal expansion perpendicular to the layers.

The observed $A_{1g}$ loses the coherence in a duration more than 20 ps, as depicted in the inset of Fig. 3b and Fig. 3d in the main text. The anharmonic effects make a major contribution to the dephasing process. In the possible multiple anharmonic decay pathways, the third-order process is assumed to play a dominant role. Here, we considered the most common pathway, Klemens channel[11], in which one optical phonon decays into two acoustic phonons (APs) with identical low-frequencies but opposite wave vectors (see the lower inset of Fig. 3d in the main text). So the anharmonic rate is correlated to the lattice temperature and follows the relationship[12]:

$$\Gamma_{\mathrm{A_{1g}}} = \Gamma_0 + \Gamma_{\mathrm{anh}} \times \left[1 + 2 \times n\left(\frac{\hbar\omega}{2}, T\right)\right], \qquad (S5)$$

where $\Gamma_{\mathrm{anh}}$ is a proportional constant, $\Gamma_0$ is the background contribution, $\omega$ is the frequency of APs, and $n$ is the thermal occupation number of APs at temperature $T$ given by:

$$n(\hbar\omega/2) = [\exp\left(\frac{\hbar\omega}{2k_{\mathrm{B}}T}\right) - 1]^{-1}, \tag{S6}$$

where $k_{\mathrm{B}}$ is the Boltzmann constant. The temperature rise $\Delta T$ in the specimen is estimated by the excited lattice temperature at equilibrium via:

$$\Delta T = \frac{f_{\mathrm{p}}\eta a}{d_{\mathrm{M}}\rho_{\mathrm{M}}C_{\mathrm{M}}}, \tag{S7}$$

where $f_{\mathrm{p}}$ is the excitation fluence, $\eta$ is the optical absorption (~40%[13]), $d_{\mathrm{M}}$ is the specimen thickness, $\rho_{\mathrm{M}}$ is the density of specimen $(\sim 7.7\ \mathrm{g\cdot cm^{-3}})$, $C_{\mathrm{M}}$ is the specific heat (~0.2189 J/g·K[13], which remains unchanged during the temperature rise given the 165 K of Debye temperature for 2H-$MoTe_2$), and $a$ (~1.82) is the correction factor for the Gaussian intensity profile of excitation spot. The $\Delta T$ is estimated as 30–150 K under various $f_{\mathrm{p}}$. Subsequently, the $n$ is calculated as 2.1–3.2. The linear relation between $\Gamma_{\mathrm{A_{1g}}}$ and $n$ is plotted in the upper inset of Fig. 3d in the main text, agreeing well with the Klemens process and testifying the anharmonic effect in the slow component of $A_{1g}$ decay.

## Supplementary Note 3: Intensity evolution of Bragg spots in UED measurements

All measured Bragg spots show similar temporal evolutions, as depicted in Fig. S6a for the results of FS specimen, indicating the dominance of lattice thermalization. The carbon support film effectively increases the radiation tolerance of CS specimen, probably by providing extra relaxation pathways for APs. Before reaching the damage threshold, the intensity drops of first two decay processes (taken at the delay times of 2 ps and 30 ps, respectively) depend linearly on the excitation fluence, as depicted in Fig. S6b, suggesting the negligible nonlinear effects. It is noteworthy that the population of longitudinal APs at M points (LAPs-M), which is the major contribution to the first intensity drop, is correlated with the phase transition of 2H to 1T' of $MoTe_2$[14].

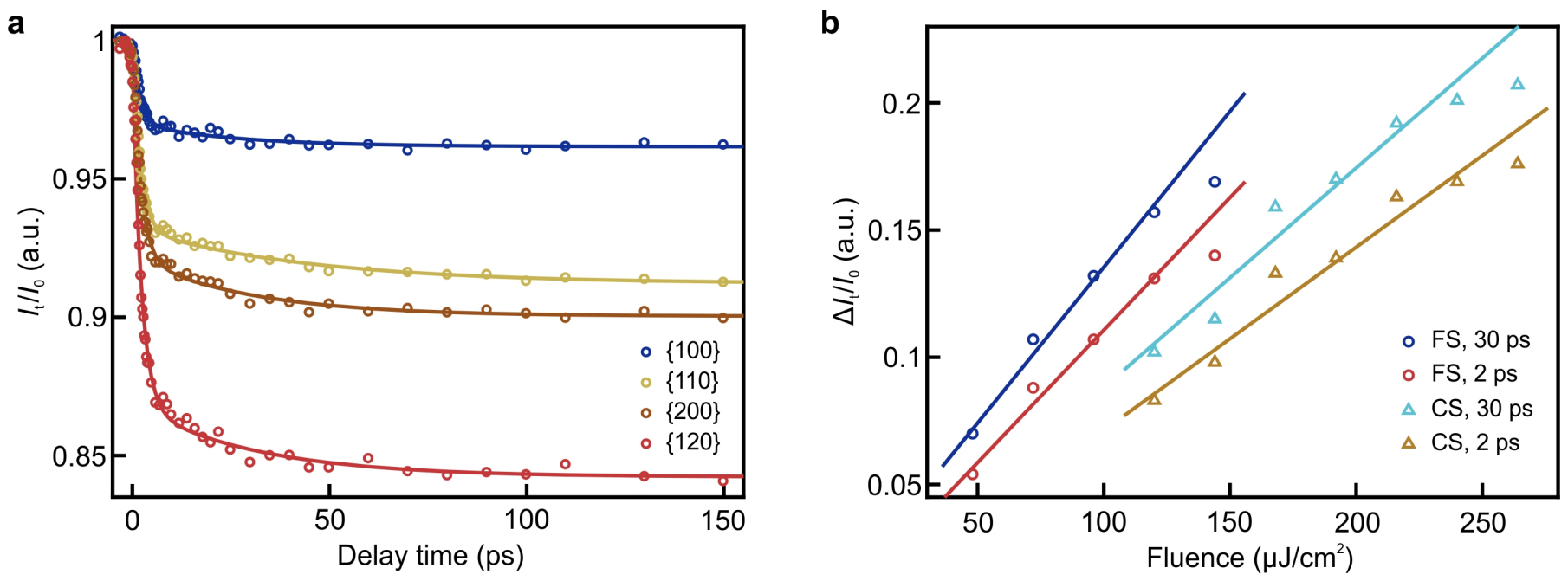

**Supplementary Figure S6: Intensity drops of measured Bragg spots and their dependence on excitation fluence.** **a**, Intensity kinetics of all 4 measured crystal plane families of the FS specimen within the short-time window, under excitation with fluence of 120 μJ/cm$^2$. Solid lines, exponential fits. **b**, Linear dependence of intensity drops of the {120} plane families on excitation fluence. The damage threshold for the FS and CS specimens are estimated as 150 and 270 μJ/cm$^2$, respectively. Solid lines, linear fits.

## Supplementary Note 4: Thermal diffusion from the specimen to the substrate

Given that the diameter of excitation spot is ~7 times bigger than that of probe spot, and the thicknesses of specimen and substrate are less than 10 nm, the probed area in UED measurements are considered uniformly excited with negligible transverse thermal diffusion. The quasi-ballistic effect[15] dominating the vertical thermal transport in both the specimen and substrate, together with the mismatches of sound velocity and phonon density of states between two materials, result in two uniform temperature distributions in each material, respectively, straddling the heterostructure interface. As the decrease of thermal energy during specimen cooling leads to heating up the substrate, the specimen temperature $T_{\mathrm{M}}$ satisfies the following equation[16]:

$$C_{\mathrm{M}}\rho_{\mathrm{M}}d_{\mathrm{M}}\frac{\partial T_{\mathrm{M}}}{\partial t} = -\sigma_{\mathrm{K}}(T_{\mathrm{M}} - T_{\mathrm{C}}), \tag{S8}$$

where $C_{\mathrm{M}}$, $\rho_{\mathrm{M}}$, and $d_{\mathrm{M}}$ are the specific heat, density, and thickness of specimen, respectively, $\sigma_{\mathrm{K}}$ is the thermal boundary conductance, $T_{\mathrm{C}}$ is the substrate temperature. Moreover, $T_{\mathrm{M}}$ and $T_{\mathrm{C}}$ fulfill the following relationship:

$$C_{\mathrm{M}}\rho_{\mathrm{M}}d_{\mathrm{M}}S\Delta T_{\mathrm{M}} = C_{\mathrm{C}}\rho_{\mathrm{C}}d_{\mathrm{C}}S\Delta T_{\mathrm{C}}, \tag{S9}$$

where $S$ is the probed area, $C_{\mathrm{c}}$, $\rho_{\mathrm{c}}$, and $d_{\mathrm{c}}$ are the specific heat, density (~1.5 $\mathrm{g\cdot cm^{-3}}$ for amorphous carbon[17]), and thickness of substrate, respectively. The temperature changes of specimen and substrate, $\Delta T_{\mathrm{M}}$ and $\Delta T_{\mathrm{C}}$, are estimated by the effective phonon temperature, $T_{\mathrm{eff}}$, which correlates to the in-plane atomic mean-squared displacements (MSD) at delay time $t$, $\langle u^2\rangle(t)$, by:

$$\langle u^2\rangle(t) = \frac{3\hbar}{2m_\mathrm{a}}\int_0^\infty \coth\left(\frac{\hbar\omega}{2k_\mathrm{B}T_\mathrm{eff}(t)}\right)\frac{F(\omega)}{\omega}\mathrm{d}\omega, \tag{S10}$$

where $m_\mathrm{a}$ is the averaged atomic mass, $\omega$ is the phonon angular frequency, and $F(\omega)$ is the normalized phonon density of states, which is obtained by the calculation of phonon dispersion (see Fig. S3) using density functional perturbation theory with Phonopy code[18,19]. The $\langle u^2\rangle(t)$ is extracted from the measured intensity $I$ of Bragg peaks with scattering vector $\boldsymbol{q}$ by:

$$\langle u^2\rangle(t) - \langle u^2\rangle(0) = -\frac{3\ln\left[\frac{I(\boldsymbol{q},t)}{I(\boldsymbol{q},0)}\right]}{|\boldsymbol{q}|^2}, \tag{S11}$$

where the index 0 denotes the time before excitation. The change of $T_\mathrm{eff}$ under excitation of 240 uJ/cm$^2$ is depicted in Fig. S7, showing the maximum specimen temperature of 672 K at 50 ps and the quasi-thermal-equilibrium with substrate at 473 K after 3 ns. Here, we ignored the temperature rise of substrate during the quick thermalization of specimen, so that we had $\Delta T_\mathrm{M}$ of ~200 K and $\Delta T_\mathrm{C}$ of ~185 K. The $C_\mathrm{c}$ is calculated through equation (S9), yielding a value of ~1 $\mathrm{J\cdot g^{-1}K^{-1}}$, which agrees well with the simulated result of 0.83 $\mathrm{J\cdot g^{-1}K^{-1}}$ at 300 K for amorphous carbon[20].

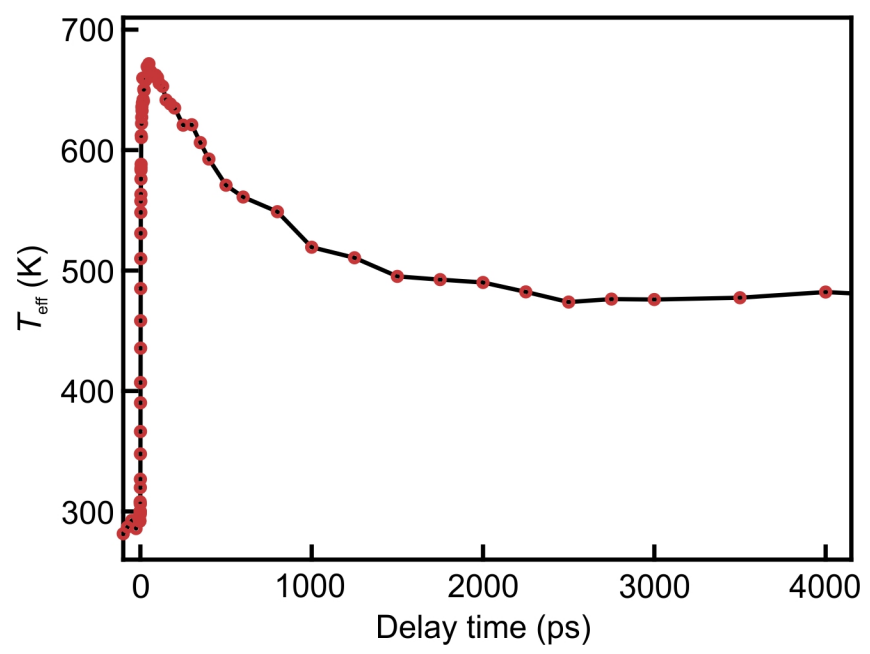

**Supplementary Figure S7: $T_\mathrm{eff}$ change of CS specimen after excitation.**

Equation (S8) can be rewritten as:

$$C_\mathrm{M}\rho_\mathrm{M}d_\mathrm{M}\frac{\partial T_\mathrm{M}}{\partial t} = -\sigma_\mathrm{K}(T_\mathrm{M} - (\frac{(T_\mathrm{M0} - T_\mathrm{M})C_\mathrm{M}\rho_\mathrm{M}d_\mathrm{M}}{C_\mathrm{c}\rho_\mathrm{c}d_\mathrm{c}} + T_\mathrm{C0})), \tag{S12}$$

where $T_\mathrm{M0}$ and $T_\mathrm{C0}$ are the initial temperatures of specimen and substrate, respectively. Equation (S12) can be reduced to:

$$C_\mathrm{M}\rho_\mathrm{M}d_\mathrm{M}\frac{\partial T_\mathrm{M}}{\partial t} = -\sigma_\mathrm{K}\left(T_\mathrm{M}\left(1 + \frac{C_\mathrm{M}\rho_\mathrm{M}d_\mathrm{M}}{C_\mathrm{c}\rho_\mathrm{c}d_\mathrm{c}}\right) + A\right), \tag{S13}$$

where $A$ is the constant term. The characteristic time $\tau$ of exponential decay[16] of $T_\mathrm{M}$ is given by:

$$\tau = \frac{C_{\mathrm{M}}\rho_{\mathrm{M}}d_{\mathrm{M}}}{\sigma_{\mathrm{K}}}\left(1 + \frac{C_{\mathrm{M}}\rho_{\mathrm{M}}d_{\mathrm{M}}}{C_{\mathrm{c}}\rho_{\mathrm{c}}d_{\mathrm{c}}}\right)^{-1}. \tag{S14}$$

Using the 670 ps of $\tau$ extracted from UED measurement, we estimated $\sigma_{\mathrm{K}}$ as ~10 $\mathrm{MW \cdot m^{-2}K^{-1}}$, which is close to the value of interface between $MoS_2$ film and $SiO_2$ substrate[21], and those of interfaces between metal films and diamond substrate[22].

## Supplementary Note 5: Fermi surface calculation

Given the measured recombination time of photo-induced carriers up to hundreds of picoseconds, the semiconductor 2H-$MoTe_2$ can be considered as a metal when the electrons stay at the conduction band[14]. The Fermi surfaces ($E_F$) of different carrier concentrations are obtained from the DFT calculation, by integrating the electron density of states, as depicted in Fig. S8a. The electron concentration in conduction band within one primitive cell, $n_{\mathrm{e}}$, under excitation with fluence of $f_{\mathrm{p}}$, is given by:

$$n_{\mathrm{e}} = \frac{f_{\mathrm{p}}\eta a s_{\mathrm{L}} c_{\mathrm{L}}}{\hbar\omega_{\mathrm{p}} d_{\mathrm{M}}}, \tag{S15}$$

where $\eta$ is the optical absorption (~40%[13]), $a$ (~1.82) is the correction factor for the Gaussian intensity profile of excitation spot, $s_{\mathrm{L}}$ is the in-plane primitive cell area, $c_{\mathrm{L}}$ is the out-of-plane lattice constant, and $\omega_{\mathrm{p}}$ is the angular frequency of excitation laser.

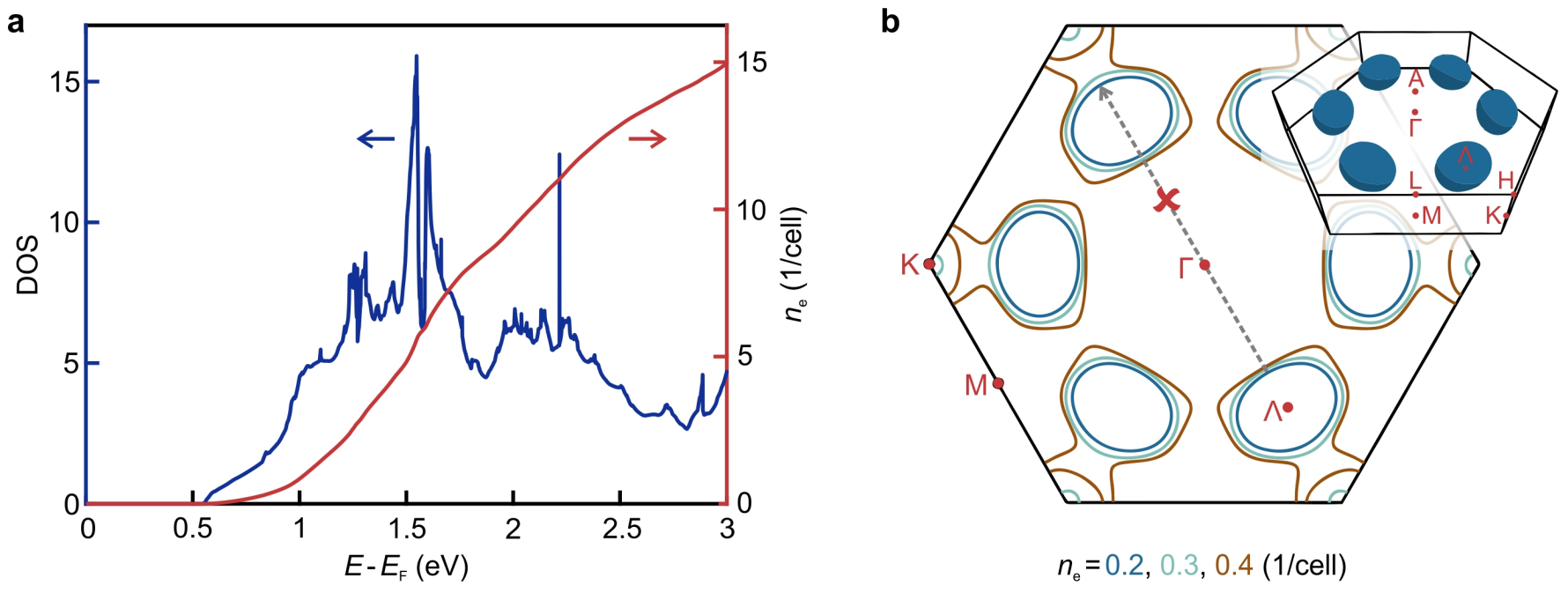

**Supplementary Figure S8: Integrating electron density of states above $E_F$ and $E_F$ contours of different electron concentrations. a**, Density of electron states (blue) and their integral (red). **b**, $E_F$ contours on (001) lattice plane for electron concentrations of $n_e = 0.2$ (blue), 0.3 (cyan), and 0.4 (brown), in the conduction band. The dashed arrow denotes the disappeared nesting vector.

The calculated $E_F$ of bulk 2H-$MoTe_2$ for the electron concentration of 0.2 e/cell, which is corresponding to an excitation of 120 μJ/cm$^2$, is depicted in the inset of Fig. S8b. The projected cross-section contours of $E_F$[23] for different electron concentrations, crossing the Γ point and perpendicular to the $c$-axis, are depicted in Fig. S8b. Due to the interlayer van der Waals force in our specimen with several atomic layers, the two parallel nesting pieces of $E_F$ in monolayer TMDCs[24] disappear, consisting with the disappearance of apex of the real part of Lindhard function in the 3D free electron gas model[25]. The lattice thermalization reflects the inelastic scattering between electrons and lattice, suggesting the significant wavevector dependence of electron-phonon coupling vertex, $\boldsymbol{g}$[26]. Hence, the strong coupling between electrons and $A_{1g}$ phonons also arises from the wavevector-dependent $\boldsymbol{g}$.

## Supplementary Note 6: Calculation of Debye-Waller effect and estimation of acoustic phonon population

The intensity of Bragg peak with scattering vector $\boldsymbol{q}$, $I(\boldsymbol{q})$, can be expressed through the Debye-Waller description[27], to quantitatively describe the atomic displacements from the equilibrium position due to thermal vibration, as

$$I(\boldsymbol{q}) = I_0(\boldsymbol{q}) \exp\left[2\left(W_0(\boldsymbol{q}) - W(\boldsymbol{q})\right)\right], \tag{S16}$$

where $I_0(\boldsymbol{q})$ is the measured intensities before excitation. The term $W(\boldsymbol{q})$ is expressed as:

$$W(\boldsymbol{q}) = \frac{\langle u^2\rangle \boldsymbol{q}^2}{2} = \frac{3\hbar^2}{2m_a k_B \theta_D}[(\frac{T}{\theta_D})^2 \int_0^{\frac{\theta_D}{T}} \frac{x}{e^x - 1} dx + \frac{1}{4}], \tag{S17}$$

where $\langle u^2\rangle$, $m_a$, $\theta_D$, and $T$ are the atomic means-square displacement, averaged atomic mass, Debye temperature, and lattice temperature, respectively. Combining equation (S16) and the left part of (S17), we have:

$$\ln(I(\boldsymbol{q})/I_0(\boldsymbol{q})) = (\langle u^2\rangle - \langle u_0^2\rangle)\boldsymbol{q}^2, \tag{S18}$$

yielding a linear relationship when the anisotropy of $\langle u^2\rangle$ is negligible, as depicted in Fig. 4b in the main text. The right part of equation (S17) is valid when the phonon density of states satisfies the Debye model[28], in which the phonon dispersion is assumed to be linear till the cutoff frequency $\omega_D$. The lattice temperature $T$ is extracted from the suppression of Bragg peaks in UED results. Taking the {120} peaks of FS specimen as an example, the obtained $T$ change and $T_{eff}$ (see Supplementary Note 4) under excitation of 120 μJ/cm$^2$ are depicted in Fig. S9a. The rise of $T$ is significantly less than $T_{eff}$ because of the very low $\theta_D$ of 165 K[13] compared to room temperature.

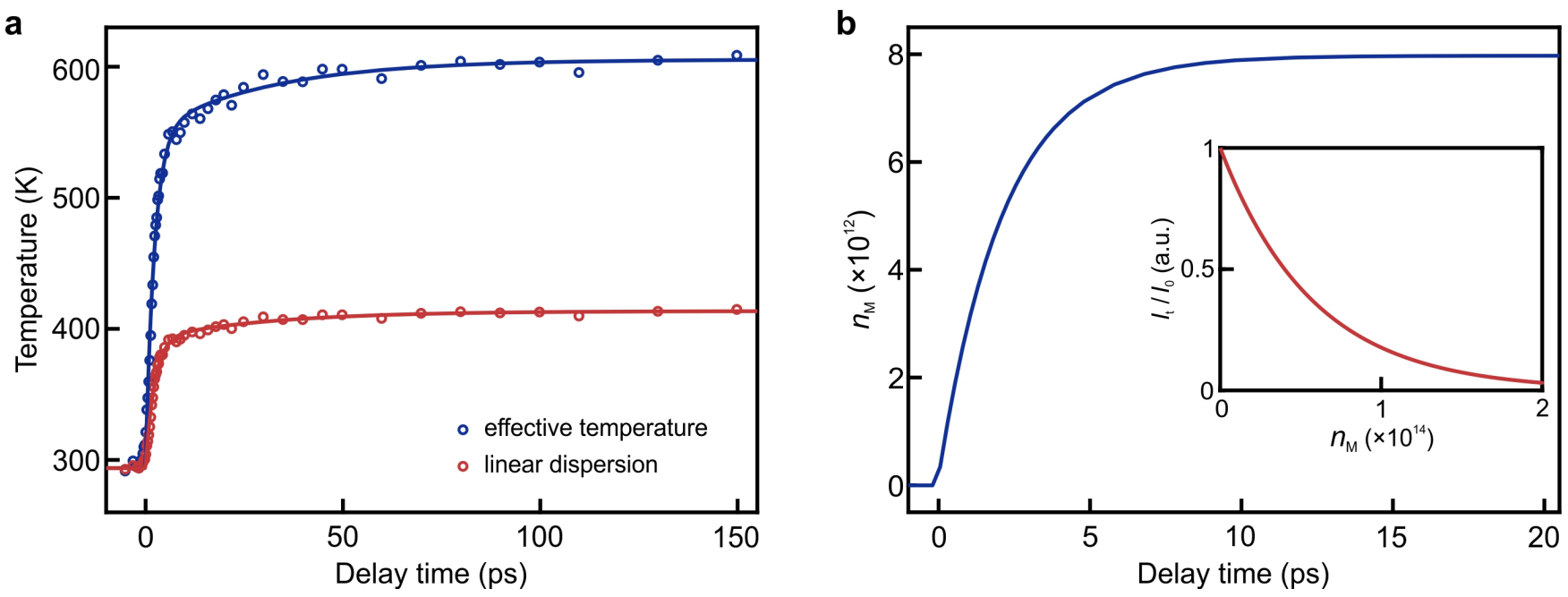

**Supplementary Figure S9: Debye-Waller effect of LAPs-M population in FS specimen. a**, Changes of lattice and effective temperatures calculated using Debye model under excitation of 120 μJ/cm$^2$. Solid lines, exponential fits. **b**, Calculated change of LAPs-M population. Inset: The dependence of $I/I_0$ on population number $n_{\mathrm{M}}$.

We now can apply a generalized description of $W(\boldsymbol{q})$, which is appropriate for the femtosecond dynamics in non-equilibrium regime[29], given by:

$$W(\boldsymbol{q}) = \frac{1}{4}\sum_{\boldsymbol{g},j} \left(\boldsymbol{q} \cdot \boldsymbol{e}_{\boldsymbol{g},j}\right)^2 \langle a_{\boldsymbol{g},j}^2 \rangle, \tag{S19}$$

where the $\boldsymbol{e}_{\boldsymbol{g},j}$ and $a_{\boldsymbol{g},j}^2$ are the polarization and mean-square amplitude of phonon, respectively, with the wave vector $\boldsymbol{g}$ in branch $j$. Taking the strongly coupling LAPs-M as the dominating mode in the first decay process of $I(\boldsymbol{q})$, the equation (S15) can be rewritten as:

$$I(t) = I_0 \exp\left(\frac{1}{2}(\boldsymbol{q} \cdot \boldsymbol{e}_{\mathrm{M}})^2 (\langle a_{\mathrm{M},0}^2 \rangle - \langle a_{\mathrm{M}}^2 \rangle)\right), \tag{S20}$$

where $I(t)$ and $I_0$ are the measured diffraction intensities after and before excitation, respectively, $\boldsymbol{q}$ is the scattering vector, $\boldsymbol{e}_{\mathrm{M}}$ is the polarization vector, and $\langle a_{\mathrm{M},t}^2 \rangle$ and $\langle a_{\mathrm{M},0}^2 \rangle$ are the mean-square amplitudes after and before excitation, respectively. $\langle a_{\mathrm{M}}^2 \rangle$ is given by:

$$\langle a_{\mathrm{M}}^2 \rangle = \frac{2(n_{\mathrm{M}} + 1/2)\hbar}{N m_a \omega_{\mathrm{M}}}, \tag{S21}$$

where $N$ is the atoms number in the probed volume, $m_a$ is the averaged atom mass, and $n_{\mathrm{M}}$ and $\omega_{\mathrm{M}}$ are the population and angular frequency of LAPs-M, respectively. The calculated change of $n_{\mathrm{M}}$ in the FS specimen is depicted in Fig. S9b. Moreover, the calculated dependence of $I/I_0$ on $n_{\mathrm{M}}$ is depicted in the inset of Fig. S9b, showing a proximate linearity when $n_{\mathrm{M}}$ is in the scale of low $10^{13}$.

In the end, the maximum populations of LAPs-M in the FS and CS specimens are estimated as ~$8\times10^{12}$ and ~$5\times10^{12}$, respectively, which are both much larger than 1, suggesting the phonon bottleneck of LAPs-M. The difference of population in two specimens also demonstrates that the substrate can effectively boost the relaxation of LAPs-M by providing extra dissipation pathways.

## Supplementary Note 7: Calculation of first-order diffuse scattering

The total intensity of electron scattering with scattering vector $\boldsymbol{q}$ and delay time $t$, $I(\boldsymbol{q},t)$, can be separated as following:

$$I(\boldsymbol{q},t) = I_0(\boldsymbol{q},t) + I_1(\boldsymbol{q},t) + \cdots, \tag{S22}$$

where $I_n$ denotes the scattered intensity of electron interacted with $n$ phonons. Specifically, $I_0$ represents the intensity of Bragg scattering, $I_1$ represents the intensity of first-order diffuse scattering. The higher-order terms are ignored in this work because of the much smaller cross sections.

In terms of $\boldsymbol{q}$, the Bragg scattering provides only limited information of specific $\boldsymbol{q}$ values. In contrast, the electron diffuse scattering covers the information of almost all the probed reciprocal space. Considering the momentum transferred to the scattered electrons, the momentum distribution of phonons is encoded in the intensity distribution of diffuse scattering, so that the intensity changes of diffuse scattering at different positions in the Brillouin zone represent the population changes of different phonon modes. Thus the evolution of diffuse scattering signal provides us an approach to assess the evolution of nonequilibrium phonons. The intensity change of first-order diffuse scattering, $\Delta I_1(\boldsymbol{q},t)$, at the M point in Brillouin zone at delay time $t$ is given by[30]:

$$\Delta I_1(\boldsymbol{q},t) \approx N_\mathrm{c} I_\mathrm{c} \sum_j \frac{\Delta n_{j,\mathrm{M}}(t)}{\omega_{j,\mathrm{M}}(t_0)} \left|F_{1j}(\boldsymbol{q},t_0)\right|^2, \tag{S23}$$

where $N_\mathrm{c}$ is the number of unit cells involved in scattering, $I_\mathrm{c}$ is the intensity of scattering from a single event, $\Delta n_{j,\mathrm{M}}(t)$ is the variation of population of the M phonon mode in branch $\boldsymbol{j}$, $\omega_{j,\mathrm{M}}(t_0)$ is the phonon frequency before excitation, and $F_{1j}(\boldsymbol{q},t_0)$ is the one-phonon structure factor which is given by[31]:

$$\left|F_{1j}(\boldsymbol{q},t_0)\right|^2 = \left|\sum_s e^{-W_s(\boldsymbol{q},t_0)} \frac{f_s(\boldsymbol{q})}{\sqrt{\mu_s}} \left(\boldsymbol{q}\cdot\boldsymbol{e}_{j,s,M}\right)\right|^2, \tag{S24}$$

where $\mu_s$, $f_s(\boldsymbol{q})$ and $\boldsymbol{e}_{j,s,M}$ are the mass, atomic scattering factor, and phonon polarization vector of atom $s$, and $W_s(\boldsymbol{q},t_0)$ is the Debye-Waller factor at time zero (before excitation) following:

$$W_s(\boldsymbol{q},t_0) = \frac{1}{4\mu_s} \sum_{j,\boldsymbol{k}} \left|a_{j,\boldsymbol{k}}(t_0)\right|^2 \left|\boldsymbol{q}\cdot\boldsymbol{e}_{j,s,\boldsymbol{k}}\right|^2, \tag{S25}$$

where $a_{j,\boldsymbol{k}}$ is the amplitude of phonon mode with wavevector $\boldsymbol{k}$ in branch $\boldsymbol{j}$, given by:

$$\left|a_{j,\boldsymbol{k}}\right|^2 = \frac{2\hbar}{N_c \omega_{j,\boldsymbol{k}}} \left(n_{j,\boldsymbol{k}} + \frac{1}{2}\right), \tag{S26}$$

where $\Delta n_{j,\boldsymbol{k}}(t)$ is the variation of population of phonon mode with wavevector $\boldsymbol{k}$.

Here, only the vibrations of 6 low-frequency (<4 THz) phonon modes, which originated from the Davydov splitting[32] of AP modes, as depicted in Fig. S10a, are considered contributing to the intensity of first-order diffuse scattering at the $M$ points. In modes I and III, there is no in-plane component of vibration of Mo atoms, while the in-plane vibration directions of Te atoms in the same layer are inverse, cancelling out each other because of the dot product in equation (S24). In modes II and V, the vibration directions of Mo and Te atoms in the adjacent layers are also inverse and counterbalanced. In mode IV, the in-plane component of vibration of Te atoms is small and counterbalanced, while the in-plane component of vibration of Mo atoms yields net value of polarization. In mode VI, the in-plane components of vibration of Mo and Te atoms all yields net values of polarization. Eventually, only the contributions from modes IV and VI are included in the calculation of first-order diffuse scattering.

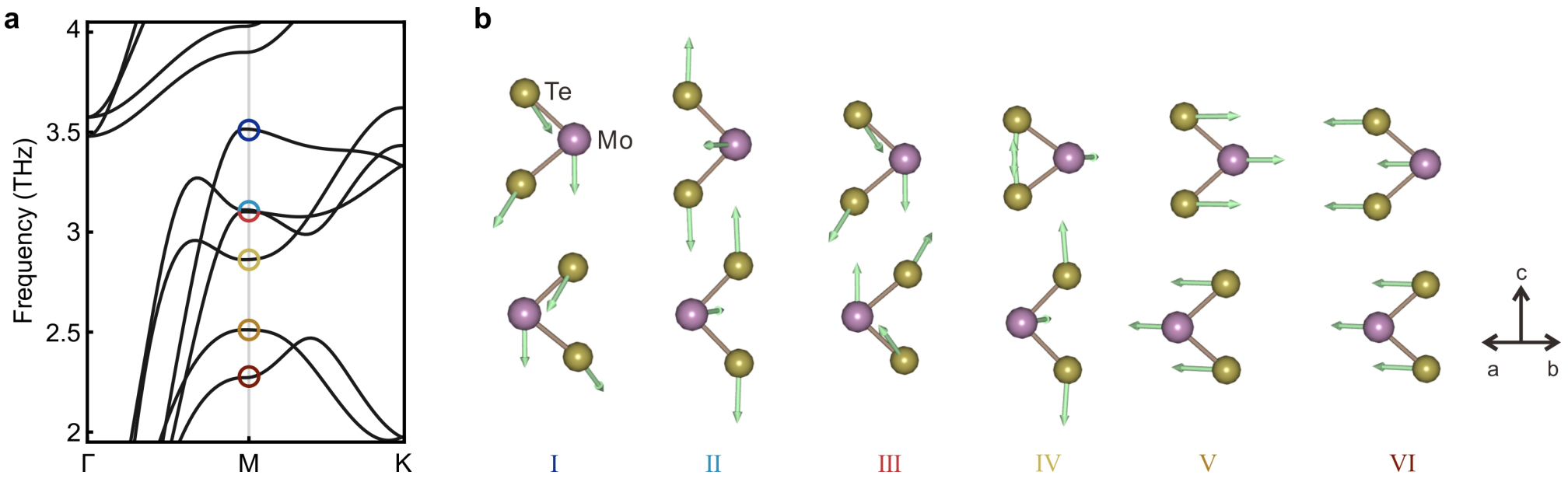

**Supplementary Figure S10: Six considered low-frequency phonon modes at M point. a**, Locations of 6 modes in the dispersion curves. Note that two locations are almost overlapped. **b**, Schematic of atomic vibrations for 6 modes in **a**.

In mode IV, i.e., the LAPs-M with frequency of ~3 THz, the vibration of Te atoms is negligible due to the direction almost perpendicular to the *a*-*b* plane, leaving only the contribution from vibration of the Mo atoms. The equation (S23) can be rewritten as:

$$\Delta I(\boldsymbol{q},t,4) = N_{\mathrm{c}} I_{\mathrm{c}} \frac{\Delta n_{4,\mathrm{M}}(t)}{\omega_{4,\mathrm{M}}(t_0)} \left| 2\mathrm{e}^{-W_{\mathrm{Mo}}(\boldsymbol{q},t_0)} \frac{f_{\mathrm{Mo}}(\boldsymbol{q})}{\sqrt{\mu_{\mathrm{Mo}}}} \left(\boldsymbol{q} \cdot \boldsymbol{e}_{4,\mathrm{Mo},\mathrm{M}}\right) \right|^2. \tag{S27}$$

The angle between the vibration direction of Mo atoms and the $\boldsymbol{q}$ is 73.9°, as depicted in Fig. S11b. Thus we have:

$$\boldsymbol{q} \cdot \boldsymbol{e}_{4,\mathrm{Mo},\mathrm{M}} = \cos 73.9° = 0.2773. \tag{S28}$$

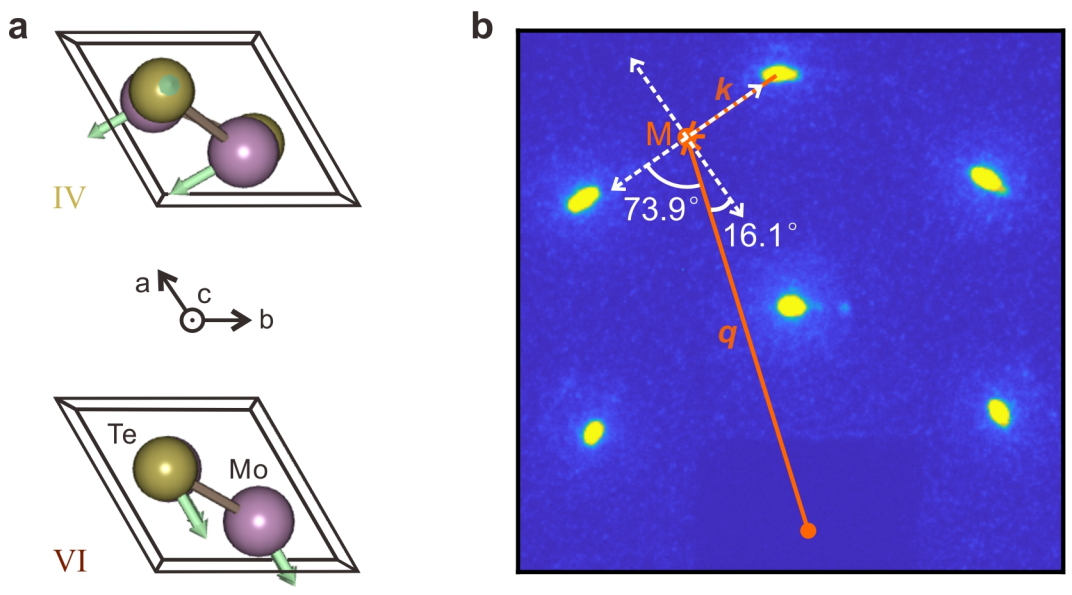

**Supplementary Figure S11: In-plane component of modes IV and VI vibration and the geometry of vibration directions.** **a**, Schematic of atomic vibrations viewing from the direction of $c$-axis. **b**, Geometry of vibration directions of Mo atoms in mode IV and Te atoms in mode VI with respect to the scattering vector $\boldsymbol{q}$, and to the reduced wavevector $\boldsymbol{k}$ which is associated to $\boldsymbol{q}$ with respect to the nearest Bragg reflection.

In mode VI, i.e., the transverse AP with frequency of ~2.1 THz, the vibrations of Mo and Te atoms need to be considered together. The equation (S23) can be rewritten as:

$$\Delta I(\boldsymbol{q},t,6) = N_c I_c \frac{\Delta n_{6,\mathrm{M}}(t)}{\omega_{6,\mathrm{M}}(t_0)} \left| 2\mathrm{e}^{-W_{\mathrm{Mo}}(\boldsymbol{q},t_0)} \frac{f_{\mathrm{Mo}}(\boldsymbol{q})}{\sqrt{\mu_{\mathrm{Mo}}}} \left(\boldsymbol{q}\cdot\boldsymbol{e}_{6,\mathrm{Mo,M}}\right) + 4\mathrm{e}^{-W_{\mathrm{Te}}(\boldsymbol{q},t_0)} \frac{f_{\mathrm{Te}}(\boldsymbol{q})}{\sqrt{\mu_{\mathrm{Te}}}} \left(\boldsymbol{q}\cdot\boldsymbol{e}_{6,\mathrm{Te,M}}\right) \right|^2. \quad \text{(S29)}$$

The angle between the vibration direction of Te atoms and the $\boldsymbol{q}$ is 16.1°, as depicted in Fig. S11b. Thus we have:

$$\boldsymbol{q}\cdot\boldsymbol{e}_{6,\mathrm{Mo,M}} = \boldsymbol{q}\cdot\boldsymbol{e}_{6,\mathrm{Te,M}} = \cos 16.1° = 0.9568. \quad \text{(S30)}$$

To compare the contributions from two modes to the scattering intensity, we need further discussions on the Debye-Waller factor $W_s$ and the atomic scattering factor $f_s$ of different atoms. From equation (S25), one can find that $W_s$ includes the contributions from all phonon modes, therefore the mass $\mu_s$ of different atom determines the value of $W_s$. But the bigger $\mu_s$ corresponds to the smaller root-mean-square displacement for the same kinetic energy of vibration, so we have $W_{\mathrm{Mo}} > W_{\mathrm{Te}}$ and $\mathrm{e}^{-W_{\mathrm{Te}}(\boldsymbol{q},t_0)} > \mathrm{e}^{-W_{\mathrm{Mo}}(\boldsymbol{q},t_0)}$.

The appropriate description of electron atomic scattering factor for most materials is a sum of fitting with Gaussian functions[33]:

$$f_s(Q) = \sum_i a_{s,i} \exp\left(-b_{s,i}Q^2\right), \quad \text{(S31)}$$

where $Q = \sin\theta/\lambda$ with $\theta$ the angle of scattering and $\lambda$ the electron wavelength, and $a_{s,i}$ and $b_{s,i}$ are the fitting parameters. Based on the neutral atom parameters provided by Peng[33], we have $f_{\mathrm{Mo}} = 3.6633$, $f_{\mathrm{Te}} = 4.2871$. Then the ratio of the contributions from modes VI and IV with same population to the intensity of first-order diffuse at M points, in the case with negligible frequency change, is estimated as following:

$$A=\frac{\frac{1}{\omega_{6,\mathrm{M}}(t_0)}\left|2\mathrm{e}^{-W_{\mathrm{Mo}}(q,t_0)}\frac{f_{\mathrm{Mo}}(\boldsymbol{q})}{\sqrt{\mu_{\mathrm{Mo}}}}\left(\boldsymbol{q}\cdot\boldsymbol{e}_{6,\mathrm{Mo,M}}\right)+4\mathrm{e}^{-W_{\mathrm{Te}}(q,t_0)}\frac{f_{\mathrm{Te}}(\boldsymbol{q})}{\sqrt{\mu_{\mathrm{Te}}}}\left(\boldsymbol{q}\cdot\boldsymbol{e}_{6,\mathrm{Te,M}}\right)\right|^2}{\frac{1}{\omega_{4,\mathrm{M}}(t_0)}\left|2\mathrm{e}^{-W_{\mathrm{Mo}}(q,t_0)}\frac{f_{\mathrm{Mo}}(\boldsymbol{q})}{\sqrt{\mu_{\mathrm{Mo}}}}\left(\boldsymbol{q}\cdot\boldsymbol{e}_{4,\mathrm{Mo,M}}\right)\right|^2} \tag{S32}$$

$$=\frac{3*\left|2\mathrm{e}^{-W_{\mathrm{Mo}}(q,t_0)}*\frac{3.6633}{\sqrt{95.94}}*0.9568+4\mathrm{e}^{-W_{\mathrm{Te}}(q,t_0)}*\frac{4.2871}{\sqrt{127.6}}*0.9568\right|^2}{2.1*\left|2\mathrm{e}^{-W_{\mathrm{Mo}}(q,t_0)}*\frac{3.6633}{\sqrt{95.94}}*0.2773\right|^2}$$

$$>\frac{3*\left|2\mathrm{e}^{-W_{\mathrm{Mo}}(q,t_0)}*\frac{3.6633}{\sqrt{95.94}}*0.9568+4e^{-W_{\mathrm{Mo}}(q,t_0)}*\frac{4.2871}{\sqrt{127.6}}*0.9568\right|^2}{2.1*\left|2\mathrm{e}^{-W_{\mathrm{Mo}}(q,t_0)}*\frac{3.6633}{\sqrt{95.94}}*0.2773\right|^2}$$

$$=\frac{3*\left|2*\frac{3.6633}{\sqrt{95.94}}*0.9568+4*\frac{4.2871}{\sqrt{127.6}}*0.9568\right|^2}{2.1*\left|2*\frac{3.6633}{\sqrt{95.94}}*0.2773\right|^2}\approx 156.1.$$

The calculation reveals that the contribution from mode VI is two orders of magnitude more than that from mode IV.

With the calculated contribution ratio, we are able to estimate the nonequilibrium population of phonon through the intensity change of diffuse scattering. Taking the FS specimen under excitation of 120 μJ/cm$^2$ as an example, the specimen reaches the quasi-thermal-equilibrium at the delay time of 150 ps, with a temperature rise of ~120 K, which is estimated though the intensity changes of Bragg peaks with the Debye-Waller effect included (see Fig. S9a). From equation (S6), the calculated populations of mode VI at room temperature (i.e., before excitation) and after thermalization are ~2.4 and ~3.6, respectively. At 150 ps, the intensity of M point increases 9% over that at room temperature, see Fig. 4d in the main text. Considering the calculated ratio and virtually only the lowest frequency mode left in the quasi-equilibrium state of phonon system, this intensity change is solely attributed to the contribution from mode VI, with a population increase of 1.2. Subsequently, we have the rates of intensity change to phonon population, e.g., 7.5% of intensity per 1 population for mode VI. At the early stage of phonon evolution, mode IV, with higher frequency, populates and decays faster, accompanied with the slowly rising mode VI, as illustrated in Fig. S12. Therefore the nonequilibrium population of mode IV can be estimated through the intensity difference between the maximum amplitude and the amplitude at 150 ps (~2.7%, see Fig. 4d in the main text). With the rate of 0.05% of intensity per 1 population for mode IV, we obtained a population of ~54 before the phonon system relaxed to the quasi-thermal-equilibrium.

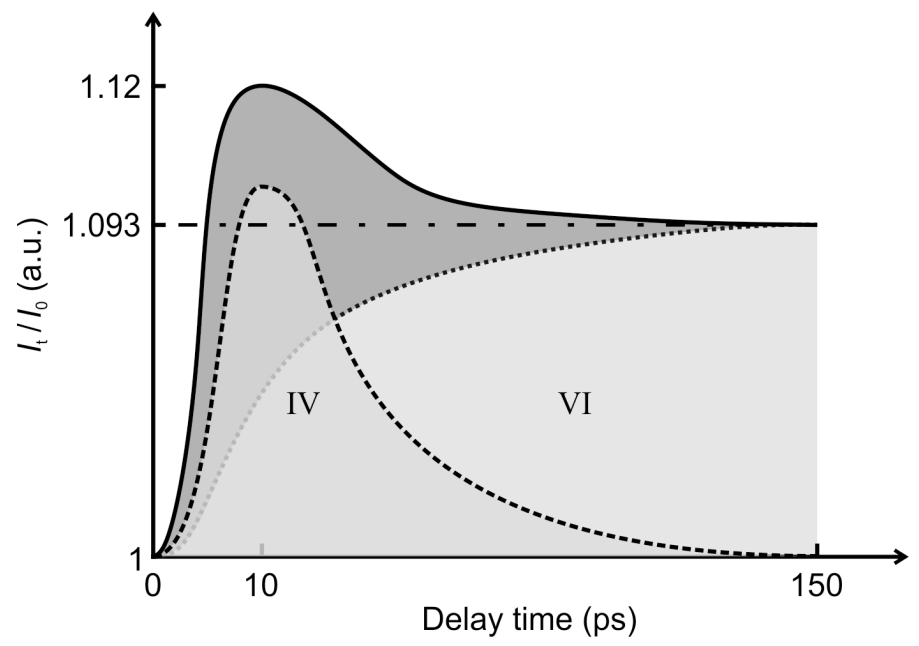

**Supplementary Figure S12: Schematic of intensity change contributed from the fast populating and decaying mode IV, and from the slowly rising mode VI.**

Note that the estimated population of mode IV is a lower bound, which is still a number indicating the occurrence of phonon bottleneck. Compared to the population estimated by the Bragg scattering with Debye-Waller effect (see Supplementary Note 6), the value by diffuse scattering is orders of magnitude smaller. In spite of the very low signal-to-noise level of diffuse scattering due to the very thin specimen thickness, the relatively small population of mode IV at nonequilibrium also suggests the intervalley scattering not limited to several specific vectors.

All phonon eigenvectors shown in this manuscript (both the main text and the Supplementary Information) are generated through the website: https://henriquemiranda.github.io/phononwebsite/phonon.html.